\documentclass[11pt,a4paper]{article}
\usepackage{jheppub}
\usepackage{tabularx}
\usepackage[flushleft]{threeparttable}
\usepackage{multirow}
\usepackage{xspace,xfrac}
\makeatletter
\usepackage{subfig}
\usepackage{tikz-feynman} 
\newcommand\footnoteref[1]{\protected@xdef\@thefnmark{\ref{#1}}\@footnotemark}
\makeatother
\usepackage{ifthen} 
\usepackage{feynmp-auto}
\newboolean{articletitles}
\setboolean{articletitles}{true} 
\usepackage{fix-cm}
\usepackage{cleveref}
\usepackage{anyfontsize}

\newcommand{\GeV}{\ensuremath{~\text{GeV}}\xspace}
\newcommand{\TeV}{\ensuremath{~\text{TeV}}\xspace}

\newcommand{\dPhi}{\ensuremath{\Delta\phi}\xspace}
\newcommand{\YZero}{\ensuremath{Y_0}\xspace}
\newcommand{\mYZero}{\ensuremath{m_{\YZero}}\xspace}

\newcommand{\mjj}{\ensuremath{m_{\mathrm{jj}}}\xspace}
\newcommand{\pT}{\ensuremath{p_{\mathrm{T}}}\xspace}
\newcommand{\pTSum}{\ensuremath{\Sigma p_{\mathrm{T}}}\xspace}

\newcommand*{\MGMCatNLOV}[1]{\textsc{MadGraph5}\_aMC@NLO~#1\xspace}
\newcommand*{\PYTHIAV}[1]{\textsc{Pythia}~#1\xspace}
\newcommand*{\DELPHES}[1]{\textsc{Delphes}~#1\xspace}

\begin{document}
\title{Enhancing di-jet resonance searches via a final-state radiation jet tagging algorithm\footnote{B.X. Liu is supported by Shenzhen Campus of the Sun Yat-sen University under project 74140-12255011, and by the Young Scientists Fund (C Class) of the National Natural Science Foundation of China (Grant No.12405122). Both Y.X. Shen and Y.S.Z. Sui are supported by College Students' Innovative Entrepreneurial Training Plan Program.}}

\author[a]{Bingxuan Liu\footnote{Email: liubx28@mail.sysu.edu.cn}}
\author[a]{Yuxuan Shen}
\author[a]{Yuanshunzi Sui}
\affiliation[a]{School of Science, Sun Yat-sen University, 66 Gongchang Road, Shenzhen, Guangdong 518107, PRC}

\abstract{ In this article, we investigate the possibility of enhancing the di-jet resonance searches by tagging the final state radiation (FSR) jet, using an event-level deep neural network. It is found that solely relying on the 4-momenta of the leading three jets allows the algorithm to achieve good discriminating power that can identify the hardest FSR jet in signal, while rejecting other soft jets. Once the invariant mass is corrected with the tagged FSR jet, the mass resolution of the signal is greatly enhanced, and the sensitivity of the search is also improved by more than 10\%. By crafting the input variables carefully, the algorithm introduces minimal mass sculpting for the background, and its applicability extends to a broad mass range. This work proves that FSR jet tagging can potentially enhance the di-jet resonance searches, suiting various stages of the physics programmes at the Large Hadron Collider (LHC) and High-Luminosity LHC (HL-LHC).}

\maketitle
\clearpage

\section{Introduction}
\label{sec:intro}

\begin{figure}[ht]
   \begin{center}
    \includegraphics[width=0.8\columnwidth]{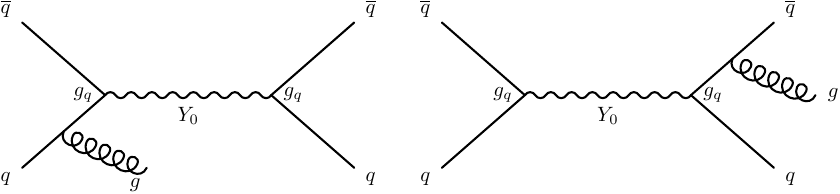}
    \caption{Feynman diagrams for a heavy \YZero particle production in $s$-channel with an ISR gluon (left) and an FSR gluon (right). The \YZero particle is a spin-0 mediator in the simplified dark matter model~\cite{dm_simp,dm_simp1,dm_simp2}.}
   \label{fig:diagrams}
   \end{center}
\end{figure}

The search for a heavy particle decaying to two jets, i.e., the di-jet search,
has a long history in collider experiments~\cite{UA1,UA2,CDF,D0,ATLASDijet,
ATLASTLA, CMSDijet,CMSDijetAD}. It is sensitive to a broad range of beyond the
standard model (BSM) theories. The heavy particle can be the mediator
connecting the standard model (SM) and BSM sectors~\cite{abcd,svj,emj}, such as 
the spin-0 mediator, \YZero, in a simplified dark matter
model~\cite{dm_simp,dm_simp1,dm_simp2}. If a heavy particle can be produced at
a hadron collider, it ought to have sizeable couplings to quarks, which
consequently gives a large enough branching ratio to di-jet final states. The
di-jet search is a natural strategy to look for such a heavy particle, and test
those relevant BSM theories. As long as there is a new particle coupled to
quarks or gluons, with a narrow decay width, the di-jet search retains its
power to any BSM models~\cite{dijetreview,atlasexo,tlareview}. Many searches
have been performed in various experiments, and the search strategy is rather
well established, given its simple event topology. Those searches usually use
the leading two jets, as they inherit most of the energy from the heavy
particle decay. However, the events rarely contain only two jets, as there can
be softer jets from initial-state radiation (ISR) and final-state radiation
(FSR), as illustrated in Figure~\ref{fig:diagrams}. At the large hadron
collider (LHC) experiments, such as ATLAS and CMS, there are also contributions
from pile-up (PU) events. To overcome the trigger threshold constraints,
experiments have also developed a search strategy that relies on an energetic
ISR jet for the triggering so that the lower mass region below 1 \TeV can be
probed without significant biases~\cite{atlasdijetisr,cmsdijetisr}. In this
case, the leading two jets are not necessarily associated with the new
particle. This work is concentrated on the mass region above 1 \TeV, where the
invariant mass of the leading two jets, \mjj, corresponds to the reconstructed
heavy particle mass. 

Although \mjj formed by the leading two jets has been proven to be effective, it is
important to thoroughly investigate the impact from the FSR. In principle,
the FSR jets should be included in the invariant mass calculation to better
reconstruct the heavy particle mass. To do so, one needs a way to identify FSR
jets while rejecting other softer jets in the events. Jets from PU are usually dealt with by the experiments using dedicated techniques~\cite{jvt}, so this study does not consider those jets.

Some previous publications have proposed the usage of ISR tagging, and constructed a
few observables~\cite{ISR2011}. It has also been discussed how the ISR
jets affect the new physics
processes~\cite{rapiditygap,jetbinning,isrmt2,qcdnewphys}. More
recently, a study explored a machine-learning-based (ML-based) technique to
classify the nature of heavy particles with the aid from the soft
jets~\cite{heavyrescolor}. However, the impact on the background and the overall
analysis is not examined extensively. It is of more importance at the current
stage of the BSM search programmes at the LHC to enhance the sensitivity, than
to distinguish the nature of new physics. In this article, we develop an FSR
jet tagging algorithm using a ML-based approach that accounts for both the signal
and background. This algorithm is constructed using basic kinematic variables
of the jets, not sensitive to details of the parton showering setups.
Meanwhile, the training procedure is designed to minimise \mjj dependence so
that it can be applied in di-jet searches using well established strategies. We show that the mass resolution of the signal can be greatly
improved, as well as the search sensitivity. In the light of high-luminosity
LHC (HL-LHC), where the integrated luminosity is expected to exceed 3000 $\mathrm{fb}^{-1}$,
the search programme may go through different phases. In the beginning, 
attention shall be paid to the discovery potential, while later the signal mass
resolution may play a more critical role after an excess is found. The method
established is capable of adopting those scenarios, owing to its flexibility.
This work identifies a promising avenue to enhance the di-jet like resonance
searches systematically, and the findings may be valuable for other hadronic
searches as well. 

The article is structured as follows, the datasets are introduced in
Section~\ref{sec:dataset}, followed by a study on the kinematic properties in
Section~\ref{sec:kinematics}; the algorithm is detailed in
Section~\ref{sec:model}, and 
Section~\ref{sec:app} discusses its applications; finally Section~\ref{sec:conclusion} summarises the
studies and offers some thoughts for future work. 

\section{Datasets}
\label{sec:dataset}

All samples used in this work are generated using
\MGMCatNLOV{2.9.18}~\cite{madgraph}, showered by \PYTHIAV{8.306}~\cite{pythia},
and reconstructed in \DELPHES{3.5.3}~\cite{delphes}. The CMS detector geometry
and performance are used for reconstruction, and the jets are clustered with a radius of $R = 0.4$, using the anti-$k_t$~\cite{antikt,catchment} algorithm.

Only the leading order process, with no additional partons, is generated with
\MGMCatNLOV, so the FSR and ISR jets are only from the parton showering step
done in \PYTHIAV. Three scenarios are considered based on the ``PartonLevel:ISR'' and ``PartonLevel:FSR'' switches~\cite{pythiacode}. The
nominal samples are showered with both switches on. Samples with either of the
two turned off are prepared to gain insights on the input variables and
validate the ISR jet labelling as discussed in Section~\ref{sec:label},
referred to as the showering control samples. Table~\ref{tab:types} summarises
those configurations.   

\begin{table}[htbp]
  \begin{center}
  \caption{Summary of the showering configurations used to produce the samples.}
\makebox[0pt]{
\renewcommand{\arraystretch}{1.2}
\begin{tabular}{|c|c|c|}
\hline
Type & PartonLevel:ISR & PartonLevel:FSR \\
\hline
nominal & on & on \\
\hline
fsr control & off & on\\
isr control & on & off\\
\hline
\end{tabular}
}
  \label{tab:types}
  \end{center}
\end{table}

\subsection{Signal}
\label{sec:signal}

The benchmark signal is a simplified dark matter model with a spin-0 mediator,
\YZero~\cite{dm_simp,dm_simp1,dm_simp2}. It has equal couplings to all types of quarks, but the
decay to a top-quark pair is not included. Model parameters are not modified to
take the recent theoretical advances or experimental constraints into account,
as the main kinematic characteristics of the model are not affected much by
those. Five \mYZero points are produced for the training step, starting from
1000 \GeV to 3000 \GeV, with a step of 500 \GeV. Each point consists of 250K
events. Four additional points are produced to test the generality of the
algorithm, starting from 3500 \GeV to 5000 \GeV, with a step of 500 \GeV.

\subsection{Background}
\label{sec:background}

The major background in di-jet resonance searches is the SM QCD multi-jet
production. As the training of the algorithm requires samples populated evenly
in the entire phase space to avoid kinematic biases, three samples sliced by
the leading jet \pT at the generation level are produced, with a cut of 450
\GeV, 900 \GeV and 1350 \GeV, respectively. The lowest \pT slice is motivated
by the usual trigger criterion applied in the inclusive di-jet analyses~\cite{ATLASDijet}.

\section{Kinematic Properties}
\label{sec:kinematics}

A heavy resonance decaying to two quarks gives rise to two energetic jets. It is usually appropriate to assume that the leading two
jets in \pT are from heavy particle decays, as long as the heavy particle mass is twice the threshold of the leading jet \pT selection. Jets from FSR are strongly
correlated with the leading two jets, while those from ISR are not. The
showering control samples are used in this section to examine these
correlations and motivate the design of the algorithm in
Section~\ref{sec:model}.  

\begin{figure}[ht]
  \begin{center}
    \includegraphics[width=0.45\columnwidth]{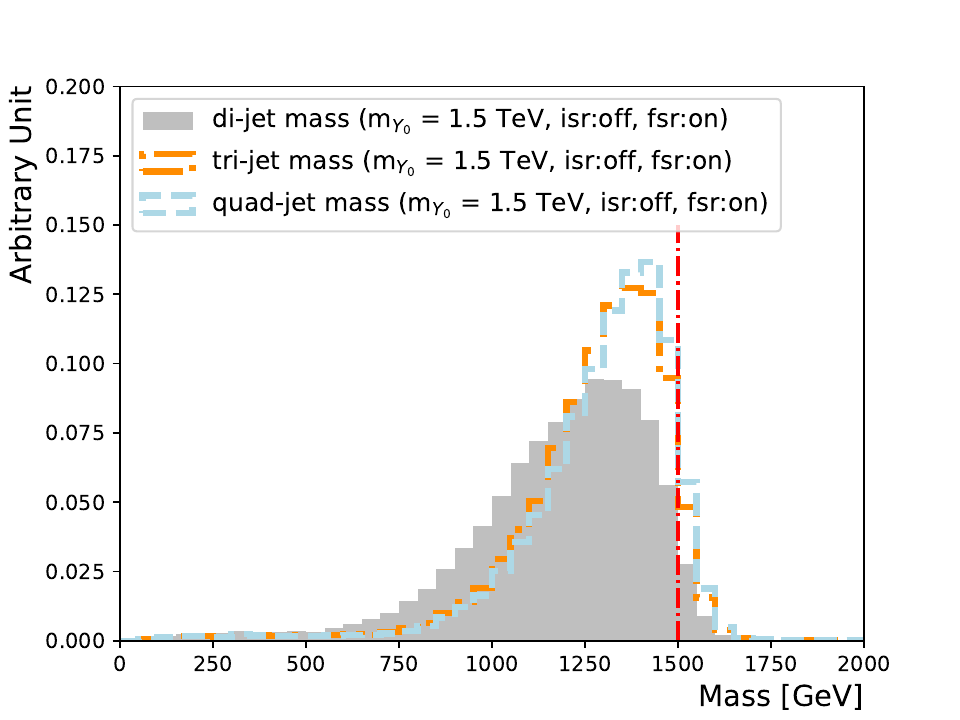}
    \includegraphics[width=0.45\columnwidth]{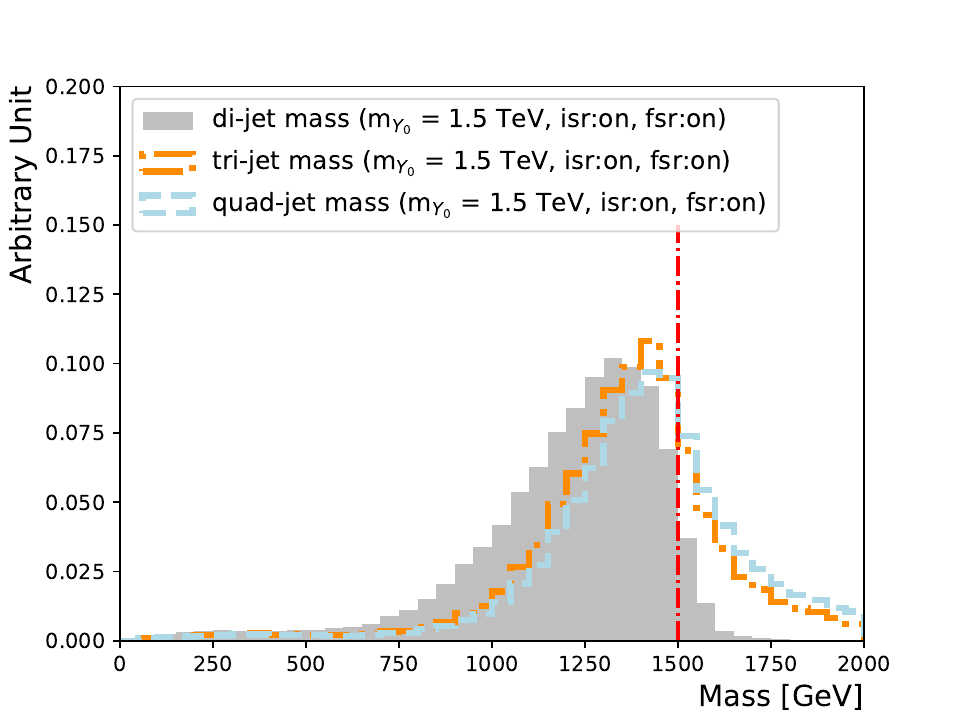}
    \caption{Comparison of the \YZero mass reconstructed using the leading two jets (shaded area), the leading three jets (dotted-dashed line) and the leading four jets (dashed line), with the ISR showering switch turned off (left) and on (right). The FSR showering switch is turned on for both. The vertical line indicates the actual \YZero mass (1.5 TeV).}
    \label{fig:fsr_mass_com}
  \end{center}
\end{figure}

An energetic FSR jet can carry away a significant amount of energy from the
heavy particle decay system, resulting in a smeared \mjj distribution. As seen in
Figure~\ref{fig:fsr_mass_com}, once including the hardest FSR jet, the mass
peak is already shifted closer to the actual \mYZero. Including additional softer FSR jets
does bring further enhancements, but it is already sufficient to showcase the
impact focusing on the hardest FSR jet. It is also obvious in
Figure~\ref{fig:fsr_mass_com} that simply including softer jets in the mass
calculation, without checking whether they are from FSR or ISR, is not a viable
strategy. It introduces a sizeable high mass tail, making the peak much
broader. 

\begin{figure}[ht]
  \begin{center}
    \includegraphics[width=0.45\columnwidth]{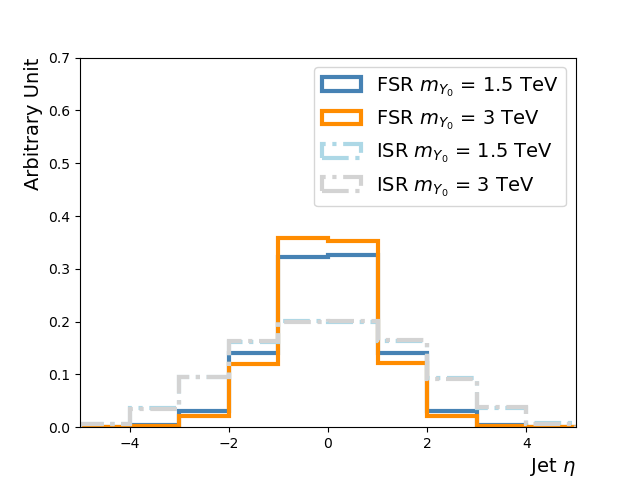} 
    \includegraphics[width=0.45\columnwidth]{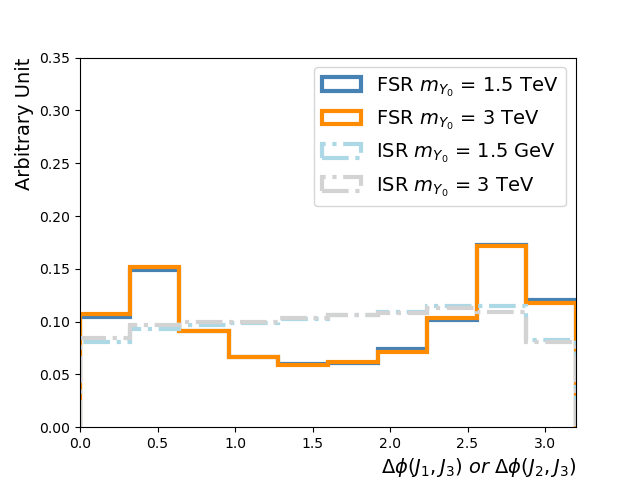}
    \includegraphics[width=0.45\columnwidth]{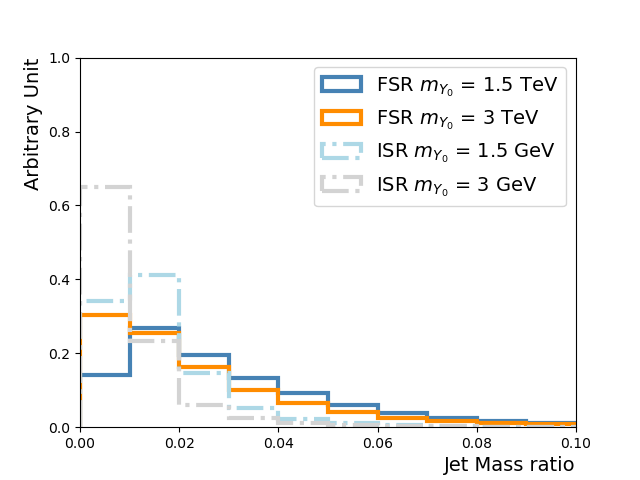}
    \includegraphics[width=0.45\columnwidth]{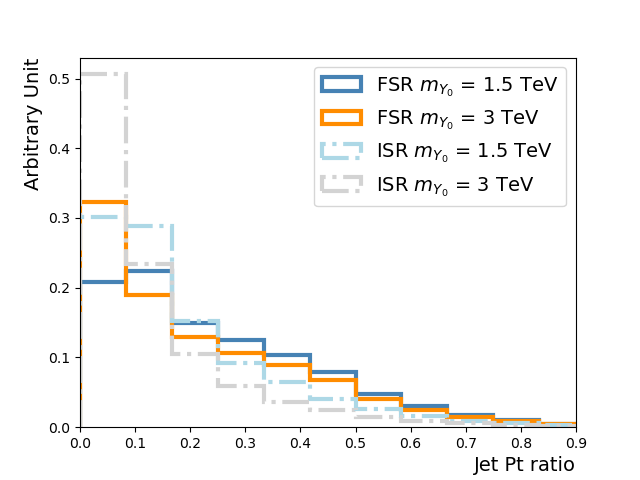}
    \caption{Selected kinematic distributions of the third jet for the \mYZero = 1.5 \TeV and 3 \TeV samples. The third jets taken from the showering control samples with the FSR/ISR showering switch turned on/off and off/on, are the FSR and ISR jets, respectively. Four quantities are shown: the third jet $\eta$ (upper left), $\Delta \phi$ between the third jet and the (sub-)leading jet (upper right),  ratio of the third jet mass (lower left) and \pT (lower right) to the leading jet \pT.}
    \label{fig:jet_com}
  \end{center}
\end{figure}

The two leading jets from a heavy \YZero particle, produced via $s$-channel, are
central and back-to-back. Since the hardest FSR jet is branched from those two
leading jets, it should be close to one of them spacially, resulting in central
$\eta$ and peaks in \dPhi w.r.t the leading two jets. The kinematic properties of
the ISR jets rely on the incoming partons, so their corresponding distributions
are wider. Figure~\ref{fig:jet_com} compares the key variables of FSR jets to those of ISR jets, using the showering control samples.    

\clearpage    

\begin{figure}[!ht]
  \begin{center}
    \includegraphics[width=0.32\columnwidth]{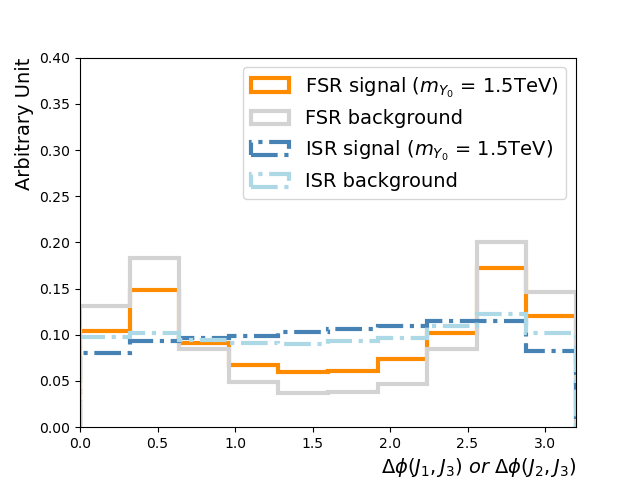}
    \includegraphics[width=0.32\columnwidth]{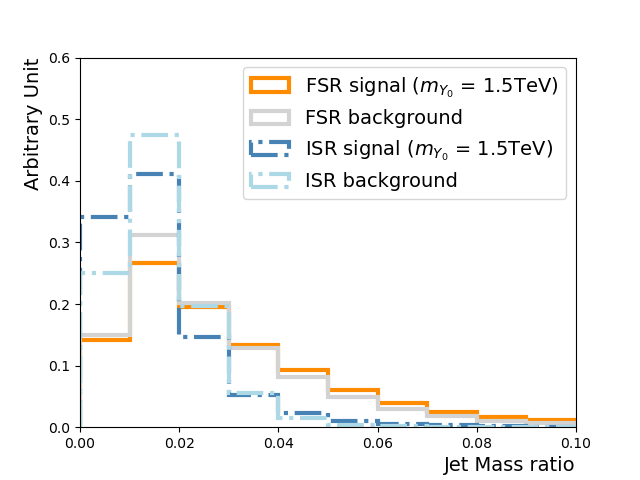}
        \includegraphics[width=0.32\columnwidth]{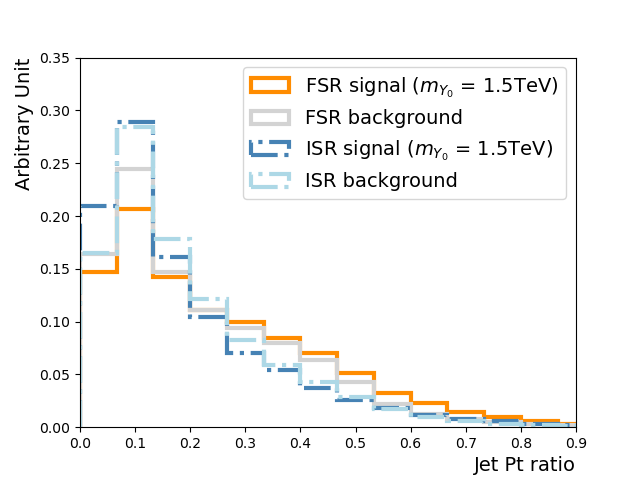}
    \caption{Selected kinematic distributions of the third jet for the \mYZero = 1.5 \TeV signal and multi-jet background. The third jets taken from the showering control samples with the FSR/ISR showering switch turned on/off and off/on, are the FSR and ISR jets, respectively. Three quantities are shown: $\Delta \phi$ between the third jet and the (sub-)leading jet (left), ratio of the third jet mass (middle) and \pT (right) to the leading jet \pT.}
    \label{fig:jet_com_signal_qcd}
  \end{center}
\end{figure}

The above observations for the signal processes still hold largely for the QCD
multi-jet, as the underlying showering process is the same, as seen in
Figure~\ref{fig:jet_com_signal_qcd}. However, high \mjj QCD multi-jet events are dominated by the $t$-channel production, and the
leading two jets are more likely to originate from gluons, compared to the
signal. Those differences allow the algorithm to distinguish FSR jets in signal from those in background. 

It is already seen that using charged particles within the jets allows us to
distinguish gluon-initiated jets from quark-initiated jets~\cite{qgtag}. The
colour connections between the radiated partons and the outgoing partons will also impact the jet constituents~\cite{colorqcd,colorbsm,rapiditygap}. Adding
lower level input features can further enhance the performance. However, doing
so makes the algorithm subject to the detailed parton shower setups and 
detector resolutions. It should be studied with great care, and we leave it for
future works.

\section{The Algorithm}
\label{sec:model}

The di-jet resonance search usually adopts a data-driven approach to estimate
the background. A classic method is to apply a functional fit to \mjj in data~\cite{UA1,UA2,CDF,D0,ATLASDijet, ATLASTLA,
CMSDijet,CMSDijetAD}. There are several new strategies proposed such as
Gaussian Process Regression~\cite{GPR1, GPR2, GPR3, GPR4}, symbolic
regression~\cite{SymbolFit} and orthonormal series~\cite{FD}. All these methods
assume the background \mjj is smooth, so significant sculpting of the \mjj
will challenge the analysis methodology. Furthermore, the di-jet searches often
try to probe a wide \mjj range without assuming the mass of the
hypothetical heavy particle. As a result, the algorithm should introduce as
minimal \mjj dependence as possible. Variables strongly correlated with 
\mjj, such as the jet \pT and mass, are not directly used in the training. As
seen in Figure~\ref{fig:jet_com} and Figure~\ref{fig:jet_com_signal_qcd},
dimensionless ratios calculated using those variables, w.r.t the leading jet
\pT, hold separation power. Those ratios are used in the training, which also
makes all the input features at a similar magnitude. Data scaling or
normalisation is found to have very minimal impact so that it is not imposed. 

Given the overwhelming multi-jet background, it is imperative to consider the
algorithm's performance there as well. If the FSR jets in the background are
tagged, the background \mjj is shifted towards higher values, which may cancel
the improvements brought to the signal \mjj resolution. Therefore, the
algorithm is designed and trained to classify four categories: ``sig-isr'',
``sig-fsr'', ``bkg-isr'' and ``bkg-fsr'', corresponding to ISR/FSR jets in the
signal/background events. The procedure to label the ISR jets is described in
the next section.

\subsection{ISR Jet Labelling}
\label{sec:label}

\begin{figure}[ht]
  \begin{center}
    \includegraphics[width=0.45\columnwidth]{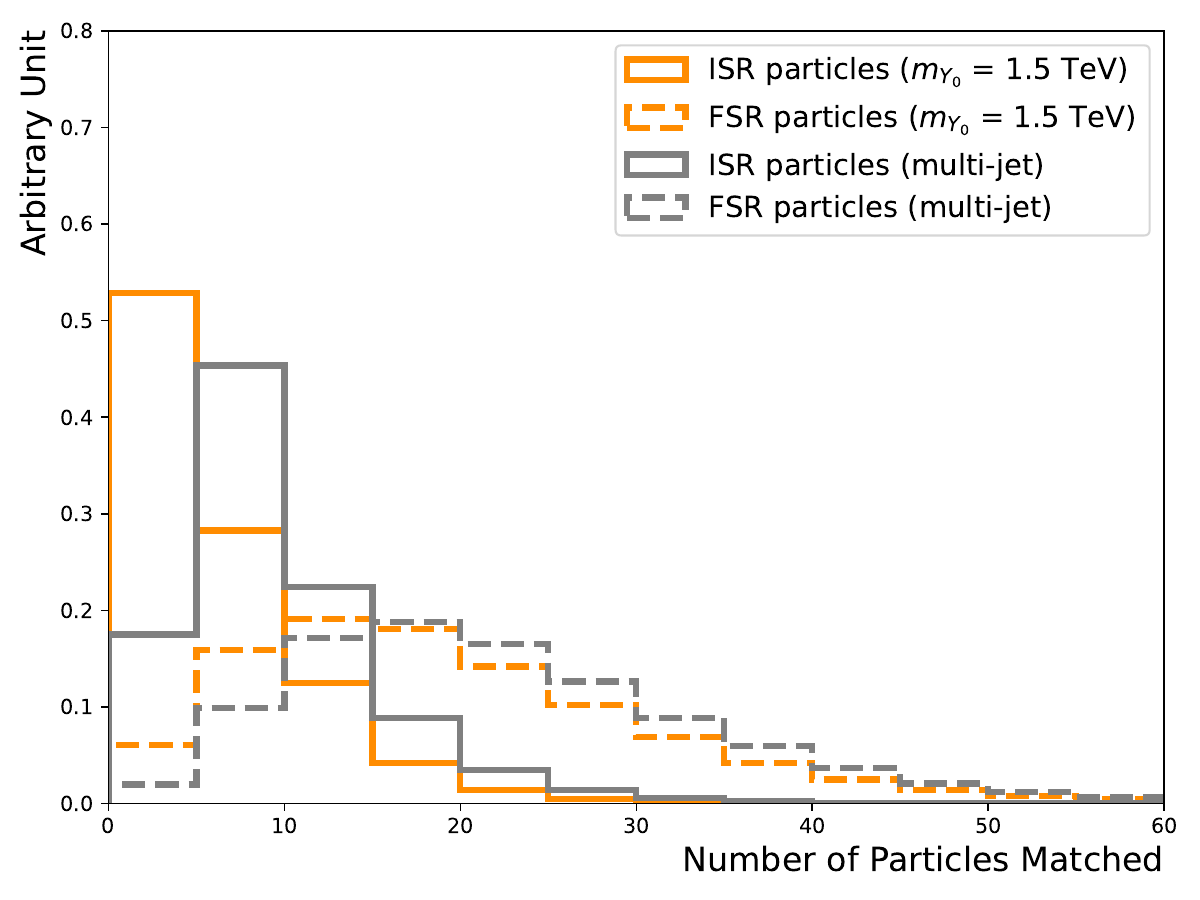}
    \includegraphics[width=0.45\columnwidth]{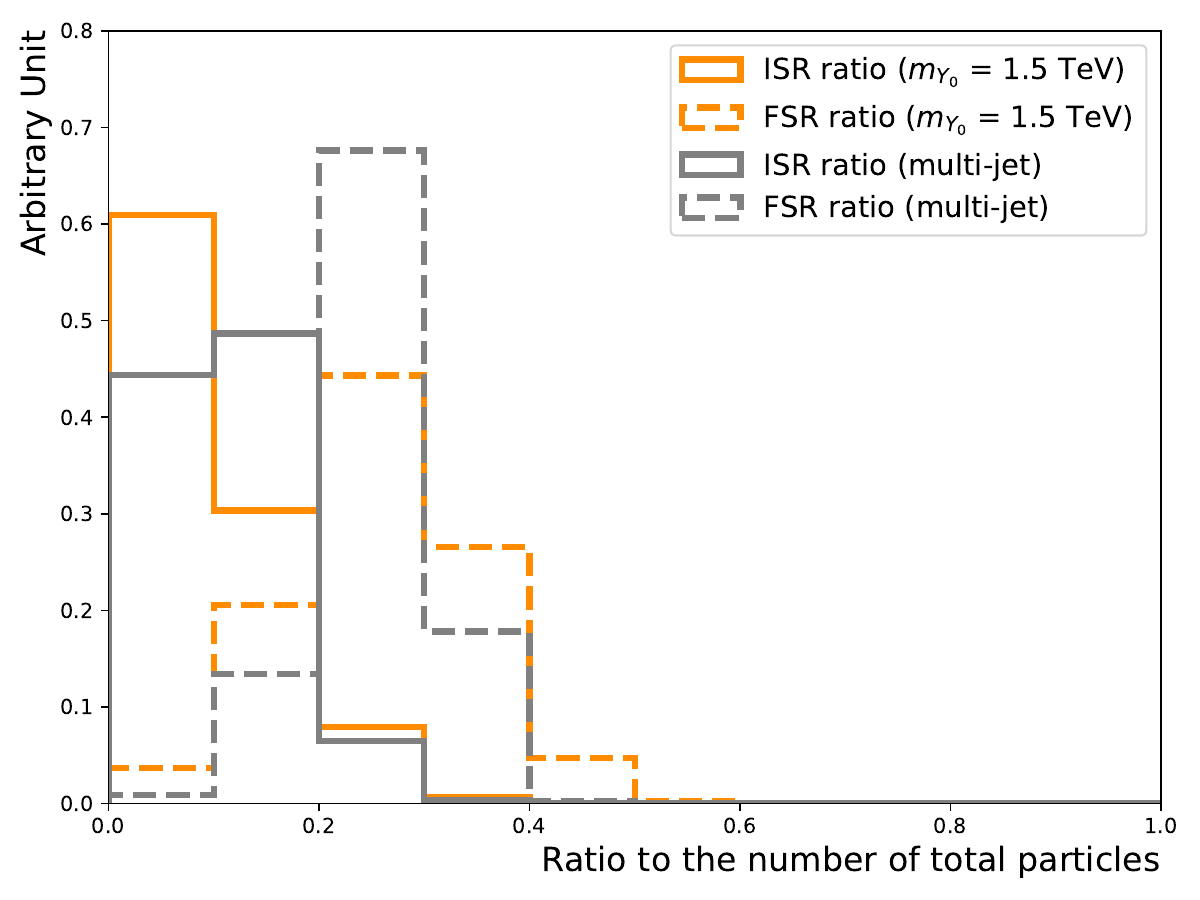}
    \caption{The number of FSR particles (dotted-dashed line) and ISR particles (solid line) associated with the third jet (left). The ratio of the number of FSR particles (dotted-dashed line) and ISR particles (solid line) to the total number of particles, including those not from ISR/FSR, associated with the third jet (right). The nominal \mYZero = 1.5 \TeV signal (dark orange) and multi-jet background (light grey) samples are used.}
    \label{fig:part_num}
  \end{center}
\end{figure}

Particles initialised by the ISR or FSR processes can be identified by the
\PYTHIAV\ status code. A status code between 41 (51) and 49 (59) means the
corresponding particle is from ISR (FSR)~\cite{pythiacode}. Those particles are matched to a
given jet by a cone with $\Delta R < 0.4$, allowing us to determine whether it
is an ISR or FSR jet. Only particles with $\pT > 0.5 \GeV$ are included, to minimise the effects from soft emissions. However, as seen in
Figure~\ref{fig:part_num}, the third jet in the event usually has both ISR and
FSR particles associated.   

\begin{figure}[ht]
  \begin{center}
    \includegraphics[width=0.45\columnwidth]{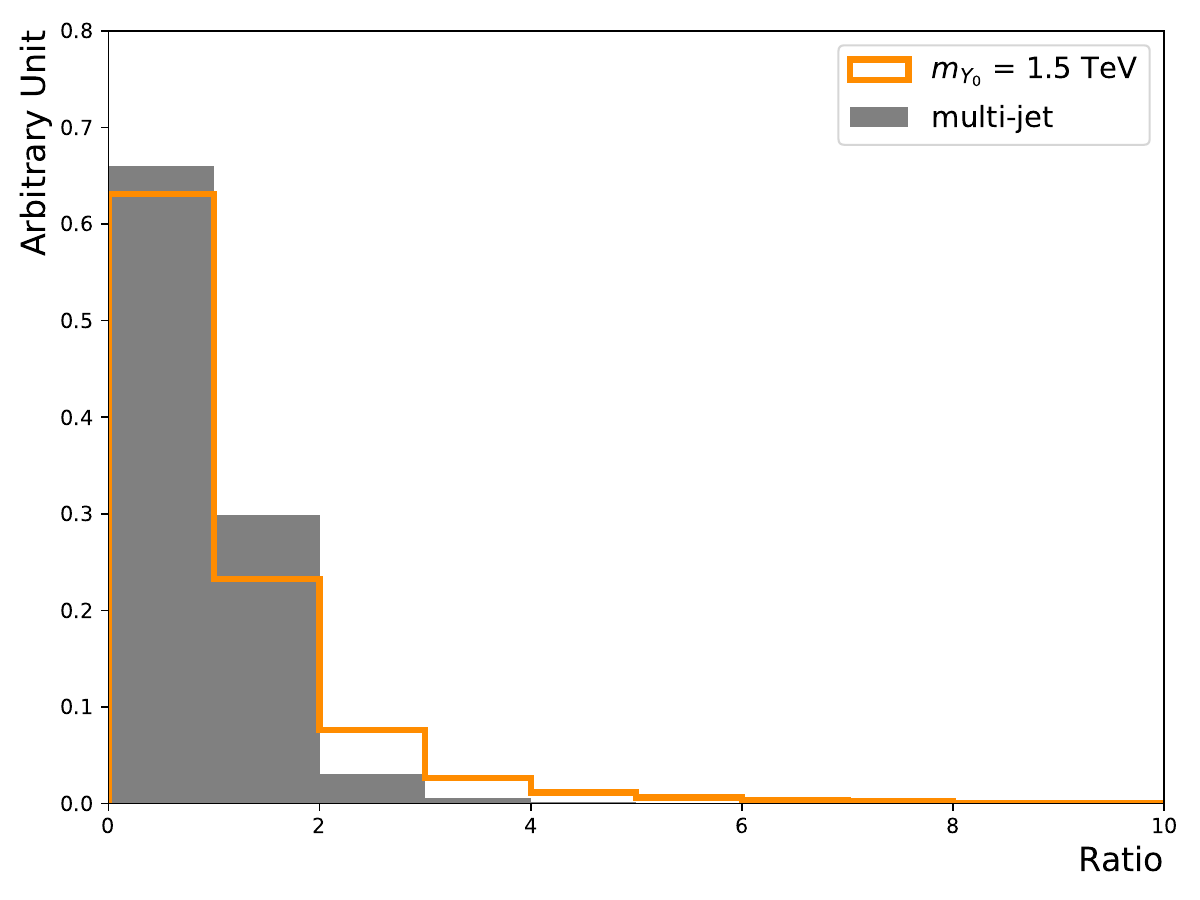}
    \caption{Ratio between the scalar sum of associated ISR particle \pT, to that of associated FSR particle \pT, for the third jet in the \mYZero = 1.5 \TeV signal (solid line) and multi-jet background (shaded area) nominal samples.}
    \label{fig:sumpt}
  \end{center}
\end{figure}

\begin{figure}[ht]
  \begin{center}
    \includegraphics[width=0.45\columnwidth]{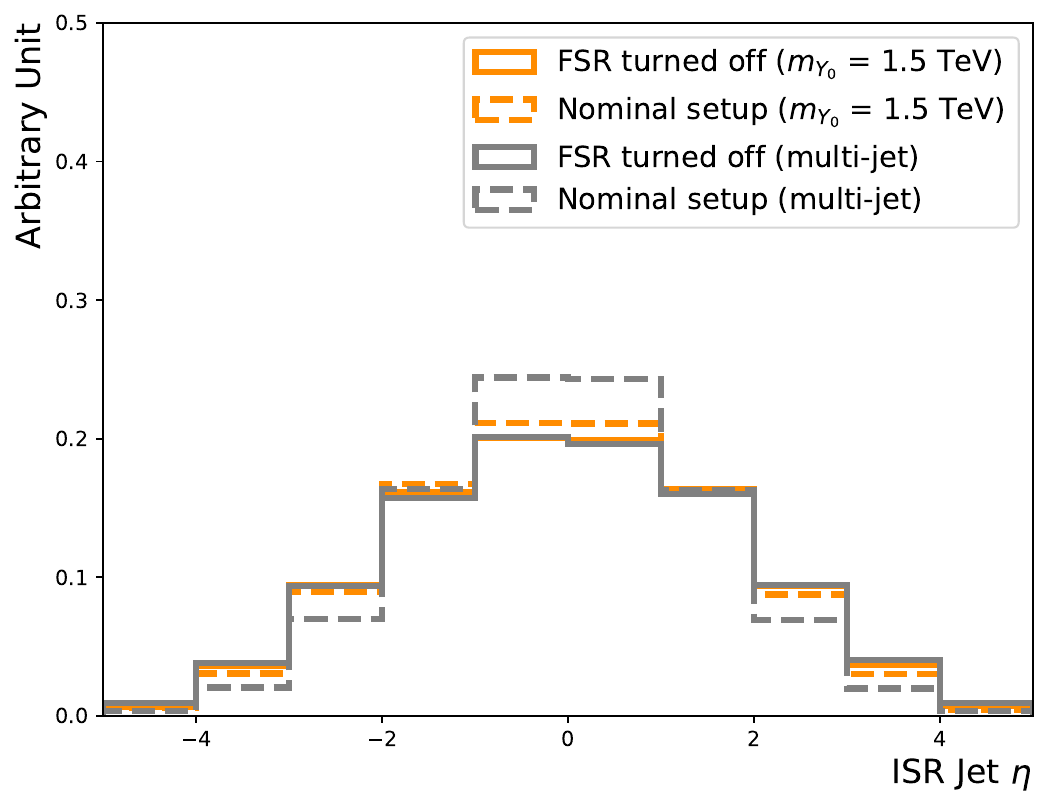}
    \includegraphics[width=0.45\columnwidth]{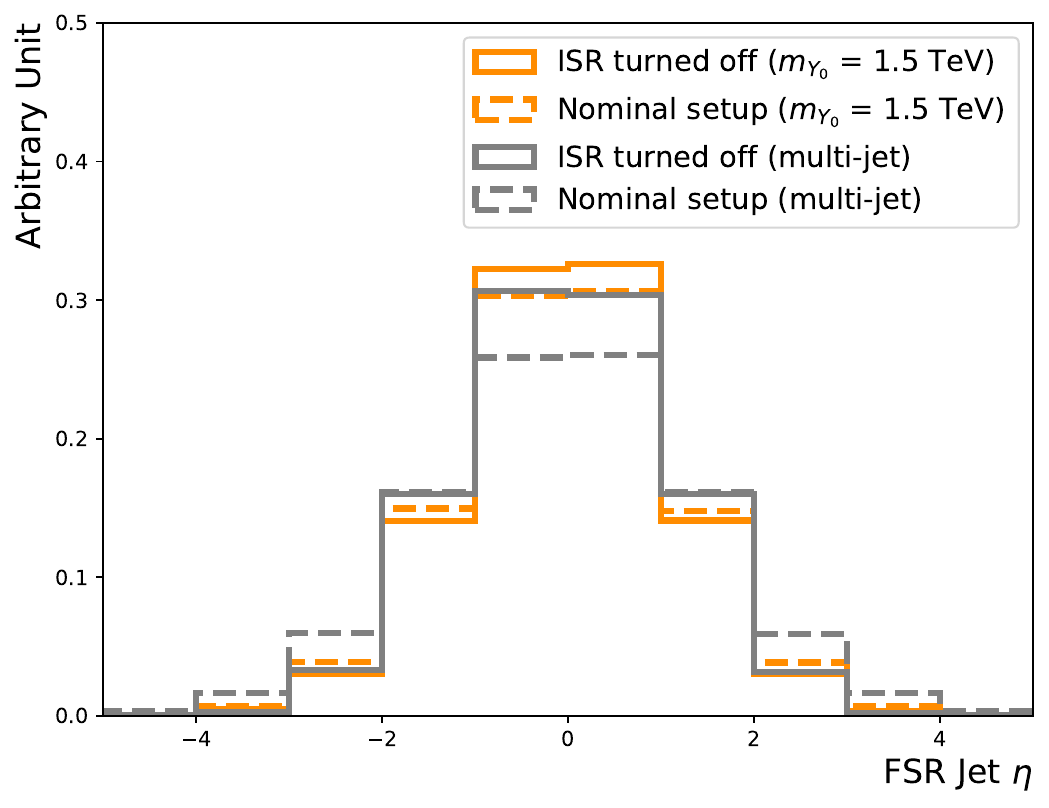}
    \caption{Comparison of the ISR jet (left) and FSR jet (right) $\eta$ (left) between those in the nominal sample labelled by the above criterion (dotted-dashed line) and those in the corresponding showering control sample (solid line). The \mYZero = 1.5 \TeV signal (dark orange) and multi-jet background (light grey) are shown.}
    \label{fig:eta_compare}
  \end{center}
\end{figure}

The scalar summation of the \pT, \pTSum, can better reflect the origin of the jets. The ratio between \pTSum of the ISR
particles to that of the FSR particles, illustrated in Figure~\ref{fig:sumpt}, is used for ISR jet labelling. Jets with
this ratio above one are taken as ISR jets. Figure~\ref{fig:eta_compare}
compares the $\eta$ distributions of the third jet obtained via this ISR
labelling method and those in the showering control samples, where reasonable
agreements are observed. The \mYZero = 1500 \GeV signal is used as an example in this section, and we observe similar behaviours for other masses as well. 

\subsection{Architecture and training}
\label{sec:arch}

The algorithm uses a simple feed-forward deep neural network, consisting of 12
input nodes, followed by four hidden layers, with 30, 60, 30 and 12 nodes,
respectively. Each node has a ReLU activation applied~\cite{relu}. A one-hot encoder is
adopted to construct the target vector with four categories. Consequently, the network has four output nodes and uses a cross-entropy loss function.

The input features include $\eta$, $\phi$ and the ratio between jet mass
and jet \pT of the leading three jets, as well as the relative fractions of the
jet momenta, summarised in Table~\ref{tab:inputs}. The background is sampled from three \pT sliced multi-jet samples,
so the events are evenly distributed across leading jet \pT. Five signal
mass points, starting from 1000 \GeV to 3000 \GeV, with a step of 500 \GeV, are
combined to populate the entire phase space. The leading jet \pT is required to
be within [450, 1750] \GeV, and the dataset is sampled to have an equal amount
of ``bkg-isr'' (``sig-isr'') and ``bkg-fsr'' (``sig-fsr'') events. The final
dataset has roughly 420k background and 460k signal events.

\begin{table}[htbp]
  \begin{center}
  \caption{Summary of the input features to train the classifier}
\makebox[0pt]{
\renewcommand{\arraystretch}{1.2}
\begin{tabular}{|c|c|}
\hline
 Type & Features \\
\hline
angular & $\eta^{\mathrm{j_1}}$, $\eta^{\mathrm{j_2}}$, $\eta^{\mathrm{j_3}}$,$\phi^{\mathrm{j_1}}$, $\phi^{\mathrm{j_2}}$, $\phi^{\mathrm{j_3}}$ \\ 
\hline
ratio & $m^{\mathrm{j_1}}/p_{\mathrm{T}}^{\mathrm{j_1}}$, $m^{\mathrm{j_2}}/p_{\mathrm{T}}^{\mathrm{j_2}}$, $m^{\mathrm{j_3}}/p_{\mathrm{T}}^{\mathrm{j_3}}$, $p_{\mathrm{T}}^{\mathrm{j_3}}/p_{\mathrm{T}}^{\mathrm{j_1}}$, $p_{\mathrm{T}}^{\mathrm{j_3}}/p_{\mathrm{T}}^{\mathrm{j_2}}$, $p_{\mathrm{T}}^{\mathrm{j_2}}/p_{\mathrm{T}}^{\mathrm{j_1}}$ \\ 
\hline
\end{tabular}
}
  \label{tab:inputs}
  \end{center}
\end{table}

The training of the algorithm takes 80\% of the dataset, with a batch size of
100. The SGD optimiser is employed~\cite{SGD}, with a learning rate of 0.05. In total, 100
epochs are carried out and the one with the best performance is selected.  

\subsection{Performance}
\label{sec:perf}

\begin{figure}[ht]
  \begin{center}
    \includegraphics[width=0.45\columnwidth]{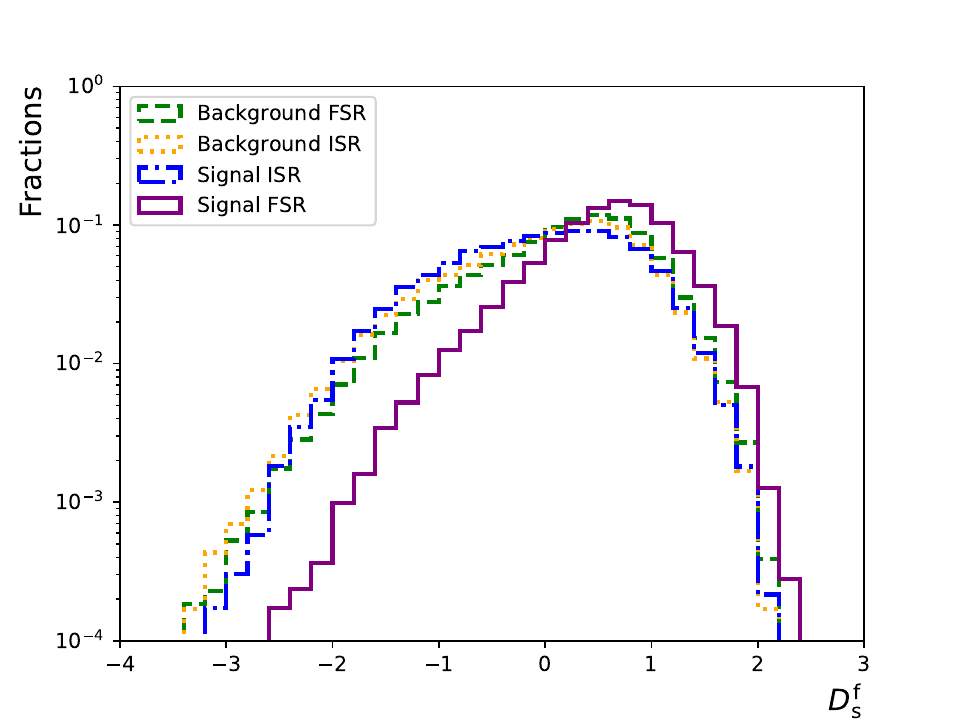}
    \includegraphics[width=0.45\columnwidth]{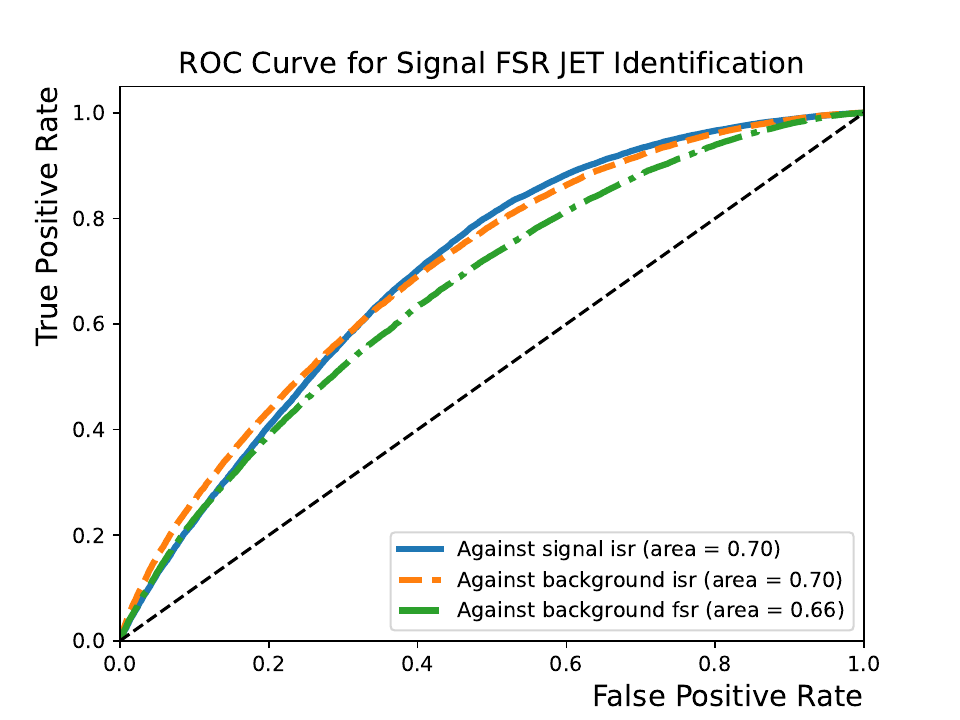}
    \caption{Left: distributions of the $D_{\mathrm{s}}^{\mathrm{f}}$ for the ``sig-isr'' (dotted-dashed line), ``sig-fsr'' (solid line), ``bkg-isr'' (dotted line) and ``bkg-fsr'' (dashed line) categories. Right: ``sig-fsr'' identification efficiency as functions of the corresponding false positive rates for the ``sig-isr'' (solid line), ``bkg-isr'' (dashed line) and ``bkg-fsr'' (dotted-dashed line) categories. They are evaluated using the test dataset that accounts for 20\% of the total combined dataset.}
    \label{fig:rocs}
  \end{center}
\end{figure}

The neural network has four output nodes, corresponding to the probabilities
for the third jet to be in those four categories: ``sig-isr'', ``sig-fsr'',
``bkg-isr'' and ``bkg-fsr''. Therefore, they are denoted as
$p_{\mathrm{s}}^{\mathrm{i}}$, $p_{\mathrm{s}}^{\mathrm{f}}$,
$p_{\mathrm{b}}^{\mathrm{i}}$ and $p_{\mathrm{b}}^{\mathrm{f}}$, respectively. A discriminating
variable can be constructed to balance the target efficiency and the false
positive rates: 

\begin{equation}
\tag{1}
 D_{\mathrm{s}}^{\mathrm{f}} = \log \frac{p_{\mathrm{s}}^{\mathrm{f}}}{(f_{\mathrm{s}}^{\mathrm{i}} \cdot p_{\mathrm{s}}^{\mathrm{i}} + f_{\mathrm{b}}^{\mathrm{f}} \cdot p_{\mathrm{b}}^{\mathrm{f}} + (1 - f_{\mathrm{s}}^{\mathrm{i}} - f_{\mathrm{b}}^{\mathrm{f}}) \cdot p_{\mathrm{b}}^{\mathrm{i}})} 
 \label{equ:score}
\end{equation}

where $f_{\mathrm{s}}^{\mathrm{i}}$ and $f_{\mathrm{b}}^{\mathrm{f}}$ are
hyperparameters that determine the relative importance. This construction is
inspired by the flavour tagging algorithms deployed by the ATLAS
experiment~\cite{GN2}. It is found that setting
$f_{\mathrm{s}}^{\mathrm{i}}$ ($f_{\mathrm{b}}^{\mathrm{f}}$) to 0.15 (0.7)
achieves similar fake rates across different processes for a given efficiency
of identifying FSR jets in signal, as seen in Figure~\ref{fig:rocs}. 

\section{Application}
\label{sec:app}

The application of the classifier can be versatile, but in this study we use it
to correct the reconstructed mass. In an event, if the third jet is identified as
coming from the ``sig-fsr'' category, the \mjj is replaced with the tri-jet invariant
mass. The impact of the classifier is three-fold. We expect to obtain improved
sensitivity, better \mjj resolutions, and good generality. A high-level discriminant, $D_{\mathrm{s}}^{\mathrm{f}}$,
is constructed in Formula~\ref{equ:score}, and can be used to select events for mass correction. The
choice of the $D_{\mathrm{s}}^{\mathrm{f}}$ threshold will affect all three metrics. In fact, in the optimal scenario, $f_{\mathrm{s}}^{\mathrm{i}}$
and $f_{\mathrm{b}}^{\mathrm{f}}$ should be tuned for each specific use-case. We perform the optimisation sequentially so that
$f_{\mathrm{s}}^{\mathrm{i}}$ ($f_{\mathrm{b}}^{\mathrm{f}}$) is not retuned,
and the cut value of $D_{\mathrm{s}}^{\mathrm{f}}$ is optimised to obtain the
best sensitivity. This workflow already gives us significant positive impact.
   
\subsection{Sensitivity}
\label{sec:sensitivity}

\begin{figure}[ht]
  \begin{center}
    \includegraphics[width=0.5\columnwidth]{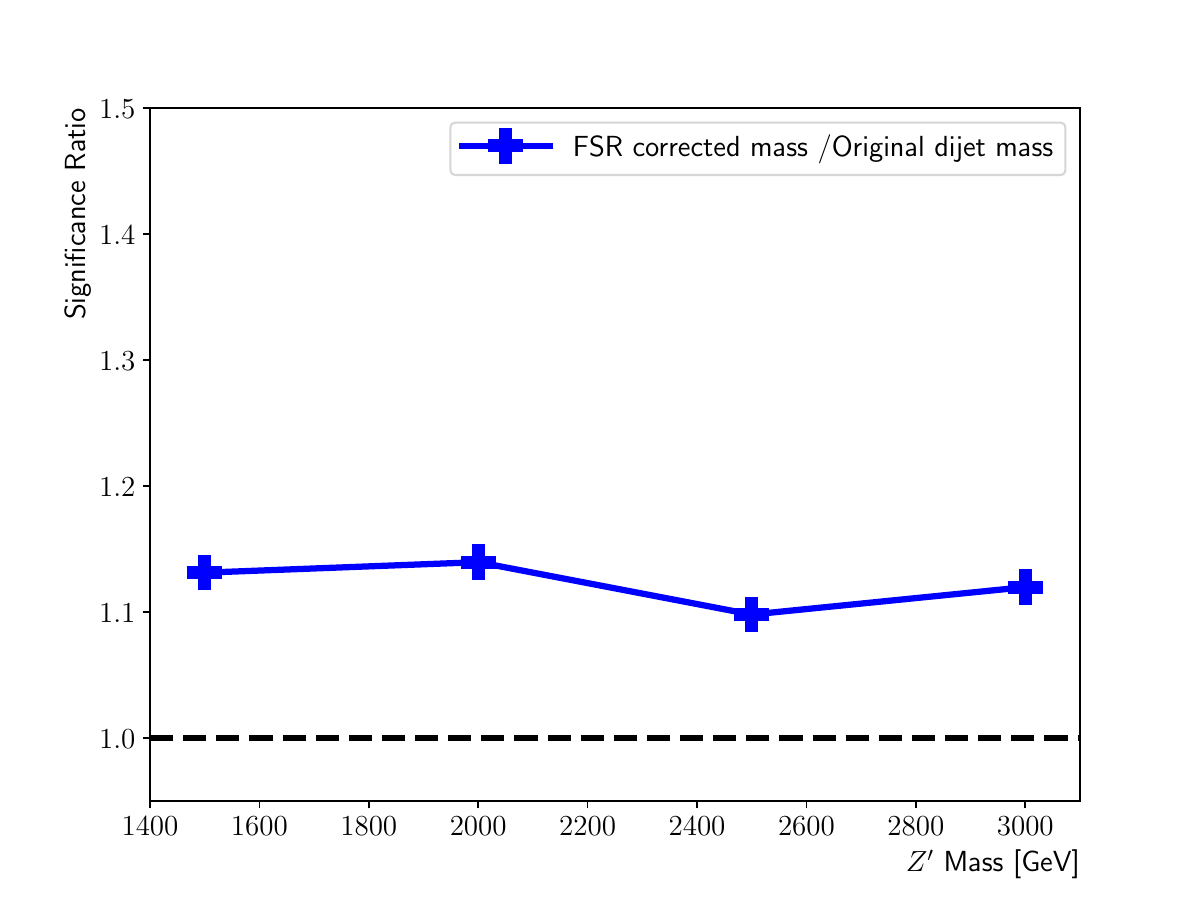}
    \caption{Summary of the ratios between $\max(\frac{\mathrm{N^i_{s}}}{\sqrt{\mathrm{N^i_B}}})$ obtained using the FSR corrected mass and that using the original di-jet mass, for the \mYZero = 1500 \GeV, 2000 \GeV, 2500 \GeV and 3000 \GeV signal points.}
    \label{fig:sen_com}
  \end{center}
\end{figure}

The sensitivity of the di-jet search can be checked by calculating the
significance, defined as $\frac{\mathrm{N}_{\mathrm{s}}}{\sqrt{\mathrm{N}_{\mathrm{b}}}}$, where
$\mathrm{N}_{\mathrm{s}}$ ($\mathrm{N}_{\mathrm{b}}$) is the number of signal (background)
events. We consider the binned \mjj, with a universal bin width of 100 \GeV, to
get the binned significance,
$\frac{\mathrm{N^i}_{\mathrm{s}}}{\sqrt{\mathrm{N^i}_{\mathrm{b}}}}$. The highest value,
$\max(\frac{\mathrm{N^i}_{\mathrm{s}}}{\sqrt{\mathrm{N^i}_{\mathrm{b}}}})$ is a good indicator of
the expected sensitivity. So the optimisation of the $D_{\mathrm{s}}^{\mathrm{f}}$
threshold is done by a coarse scan to maximise 
$\max(\frac{\mathrm{N^i_{s}}}{\sqrt{\mathrm{N^i_{b}}}})$, considering the
nominal test samples. It is found that $D_{\mathrm{s}}^{\mathrm{f}} = -0.11$ gives the largest
performance gain, where
$\max(\frac{\mathrm{N^i{s}}}{\sqrt{\mathrm{N^i_{b}}}})$ is improved by 10-14\%
for the mass range between 1.5 and 3 \TeV, as summarised in
Figure~\ref{fig:sen_com}. In the analysis, the signal strength is usually extracted via a signal + background fit, where the signal shape also plays a pivotal role. So the actual sensitivity gain will depend on the fit model adopted. 

\subsection{Mass Resolution}
\label{sec:mass_res}

\begin{figure}[ht]
  \begin{center}
    \includegraphics[width=0.45\columnwidth]{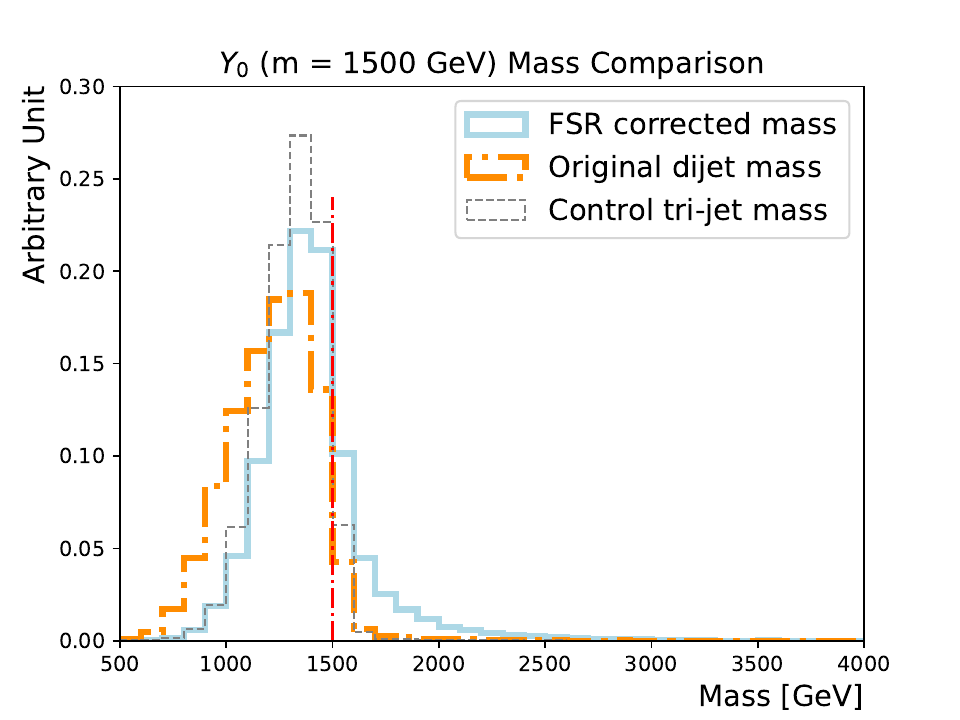}
    \includegraphics[width=0.45\columnwidth]{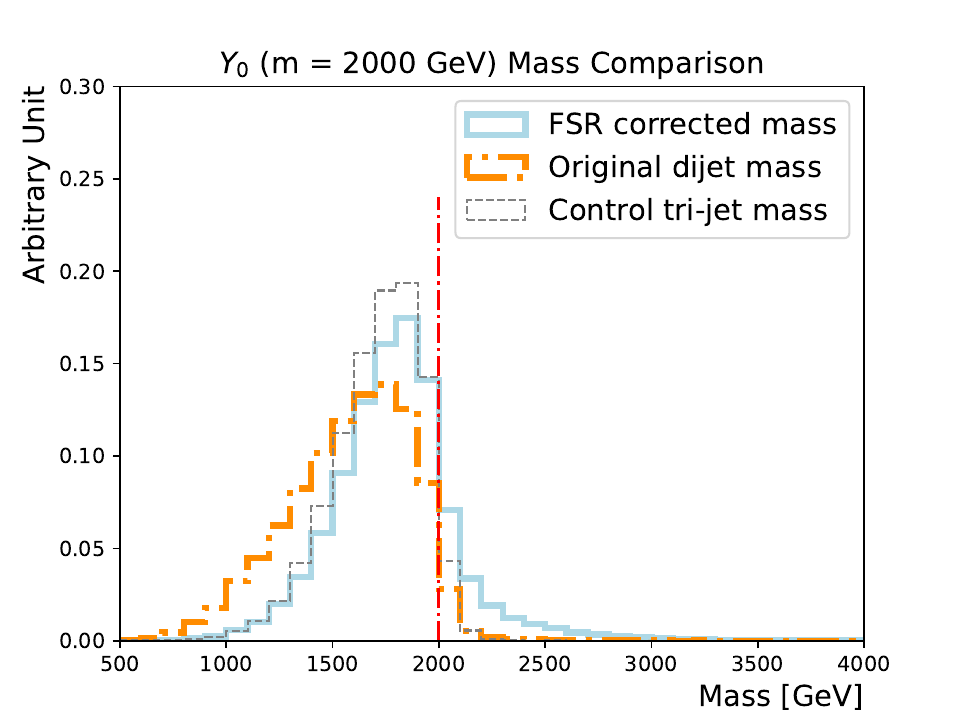}
    \includegraphics[width=0.45\columnwidth]{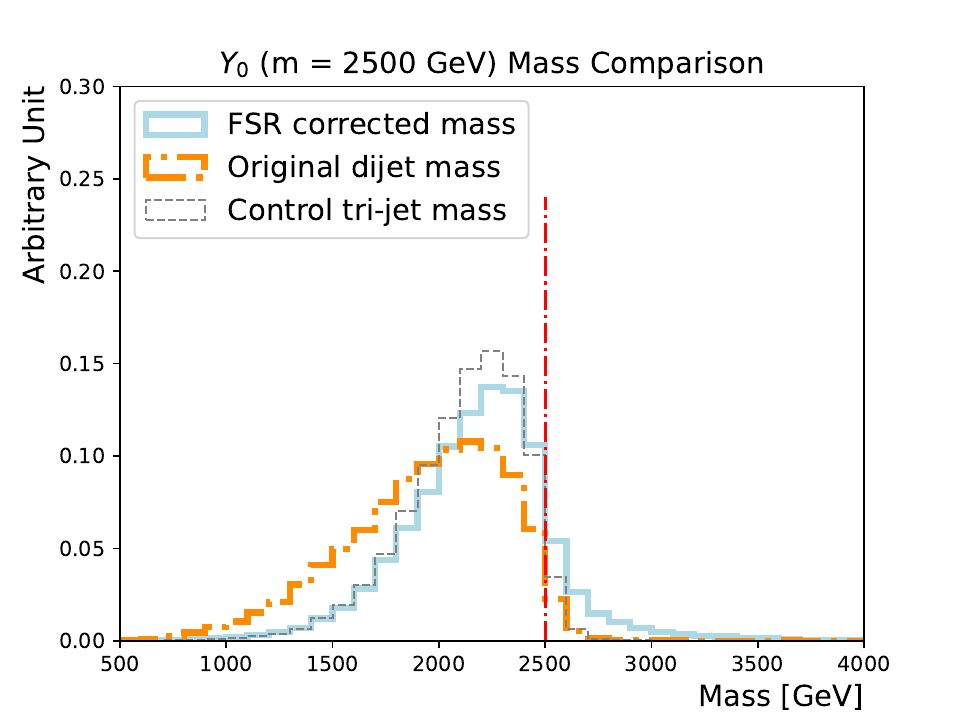}
    \includegraphics[width=0.45\columnwidth]{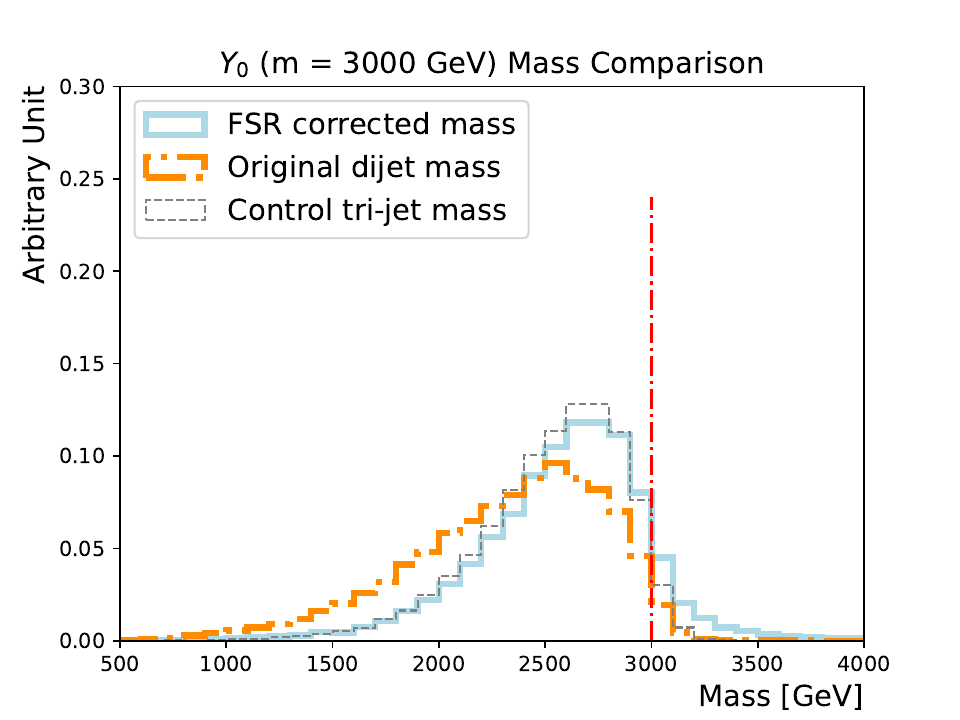}
    \caption{Comparisons of the FSR corrected masses (solid line) and the original di-jet masses (dotted-dashed line), for the \mYZero = 1500 \GeV (upper left), 2000 \GeV (upper right), 2500 \GeV (lower left) and 3000 \GeV (lower right) nominal signal samples. The tri-jet mass calculated using the showering control samples with the FSR/ISR showering switch turned on/off is added as a reference (dashed line). The actual \mYZero is indicated by the vertical line.}
    \label{fig:mass_com_sig}
  \end{center}
\end{figure}

\begin{figure}[ht]
  \begin{center}
    \includegraphics[width=0.5\columnwidth]{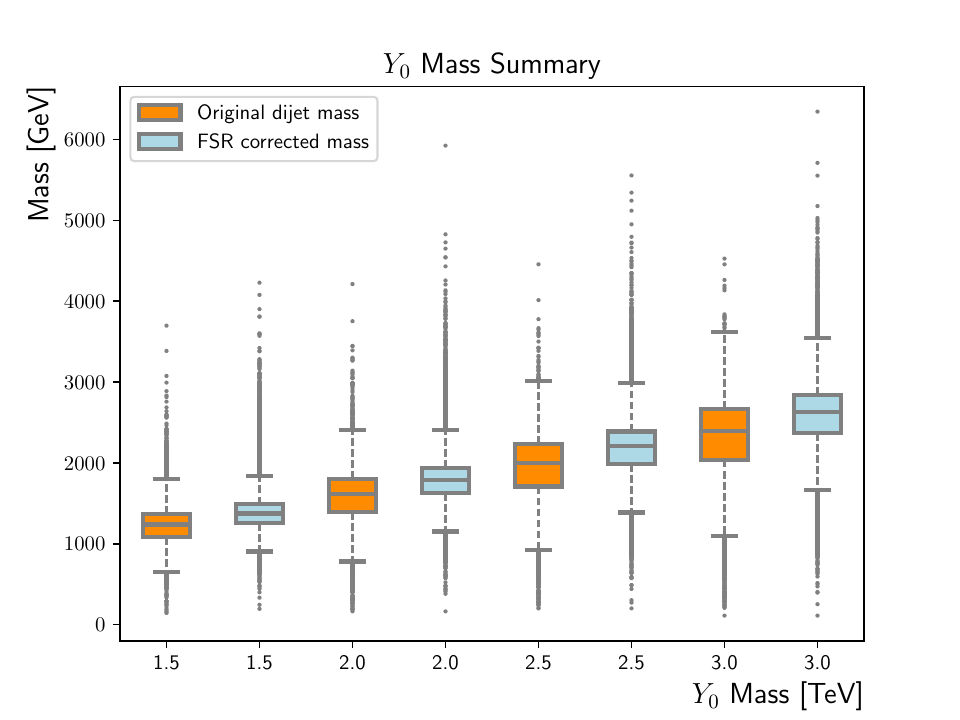}
    \caption{Summary of the FSR corrected mass distributions (light blue) and the original di-jet masses (dark orange), for \mYZero = 1500 \GeV, 2000 \GeV, 2500 \GeV and 3000 \GeV. The upper/lower boundary of the box indicates the 75\%/25\% percentile, while the centre line represents the median. The upper/lower error bar corresponds to the 95\%/5\% percentile.}
    \label{fig:mass_box_sig}
  \end{center}
\end{figure}

Applying the above $D_{\mathrm{s}}^{\mathrm{f}}$ threshold, the reconstructed mass shows
obviously a narrower peak near the actual \mYZero, as displayed in
Figure~\ref{fig:mass_com_sig}. The mild tail above \mYZero comes from ISR jets
mis-identified as FSR jets. Figure~\ref{fig:mass_box_sig} compares the impact
on reconstructed mass, showing the median is shifted towards \mYZero and
the spread becomes smaller. 

\begin{figure}[ht]
  \begin{center}
    \includegraphics[width=0.5\columnwidth]{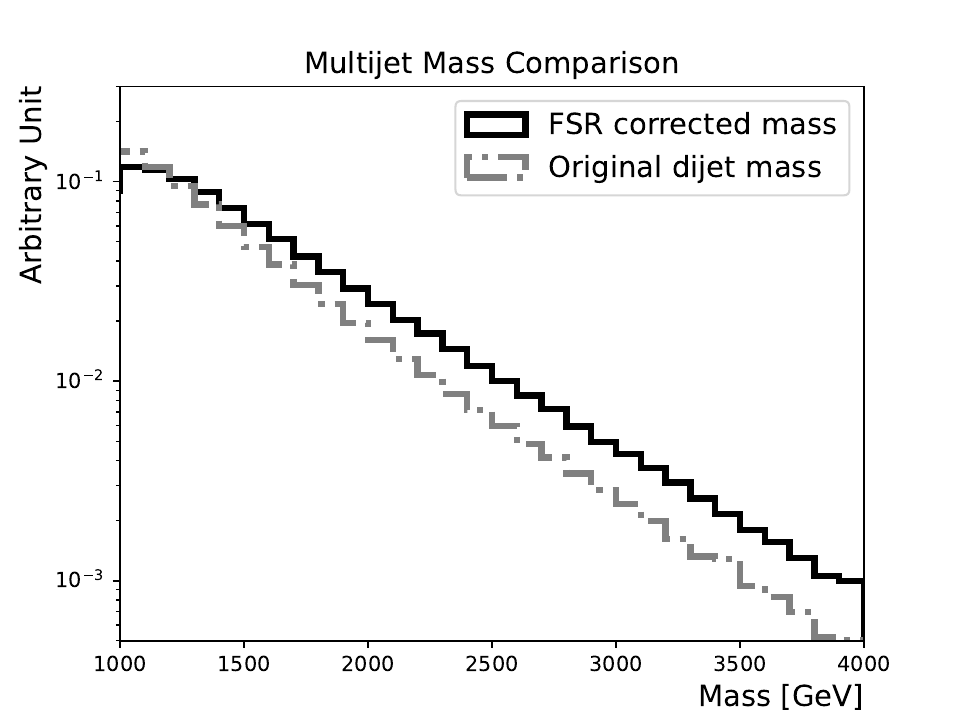}
    \caption{Comparison of the FSR corrected masses (solid line) and the original di-jet masses (dotted-dashed line), for the multi-jet background.}
    \label{fig:mass_com_bkg}
  \end{center}
\end{figure}

Since the background estimation methods applied in di-jet resonance searches
require the background mass to be smoothly falling, it is pivotal to ensure the
algorithm does not have significant mass sculpting.
Figure~\ref{fig:mass_com_bkg} overlays the original \mjj and the corrected mass
for the background process, and both have a smoothly falling behaviour.

The method developed has great flexibilities that allow us to re-optimise the
model performance for various scenarios. Two examples are given in
Appendix~\ref{app:app}.

\clearpage

\subsection{Discussion}
\label{sec:dis}

We acknowledge that the above sensitivity assessment metric is rather simple, as
in reality one needs to perform the actual background estimation. It is expected
that in a functional fit, the width of the signal plays a critical role as well 
as the signal-to-background ratio. In the future, if advanced background
modelling methods do not need to assume the background is smoothly falling any
more, this methodology has the flexibility to be re-optimised for such a scenario.

The hypothetical \YZero particle in the benchmark BSM model has democratic
couplings to quarks, and it is not coupled to gluons. Therefore, the only
production channel is via $q\overline{q}$ fusion, and the final state is
$q\overline{q}$ as well. In other theory frameworks, such as the excited
quarks~\cite{excitedquark1,excitedquark2}, extra
dimensions~\cite{extradimension} and quantum black holes~\cite{qbh}, the new
particles can couple to both quarks and gluons. So the production (decay)
channels also include $gg$, $gq$ and $g\overline{q}$. Since the algorithm is
trained utilising jet mass, a variable correlated with the jet origin,
implicitly, its performance is sensitive to the relative fraction between
quark-initiated and gluon-initiated jets in the events. The algorithm trained
with \YZero is likely not optimal for those alternative cases. Once
including lower level inputs such as the tracks within the jets, the algorithm
will have stronger BSM model dependence, which has to be carefully thought of. Another factor omitted in this study is the interference, either between the ISR and ISR processes, or between the signal and multi-jet processes. Such effects have been studied in the past, but they have not been considered in the experiments yet~\cite{inter1, inter2, inter3}. They can potentially become more significant at HL-LHC, so we should pay more attention in future investigations.

The jet clustering method used is anti-$k_t$~\cite{antikt,catchment} with a
radius of $R = 0.4$, which is the current standard choice for small-radius jets
in both the CMS and ATLAS experiments. It is found that in certain BSM
models such as dark QCD, using a larger jet radius can better reconstruct
the heavy particle mass~\cite{cmssvj, bingsvj}. Furthermore, a large jet radius
may mitigate the energy loss due to the FSR, as more final state particles from
the heavy particle decay will be clustered. So if a different jet clustering
radius is applied, the algorithm has to be re-trained, and very likely the
conclusion will change.   

\section{Conclusion}
\label{sec:conclusion}

A classifier is developed to identify FSR jets in \YZero events while
rejecting both the ISR jets in \YZero and FSR/ISR jets in multi-jet
background. The identified signal FSR jet is used to correct the reconstructed
mass, which improves the mass resolution and the sensitivity. It uses only the
variables that are not sensitive to \mjj, so an improvement of 12-20\% in sensitivity is observed across a large mass region. 

It is remarkable that only using the 4-momenta of the leading three jets 
already ensures promising performance. The classifier can achieve a 40\% signal
FSR jet identification efficiency while the fake rate of the other sources is
at $\sim 20$\% level, by constructing the discriminant accordingly. The
classifier is flexible to be adapted for different goals, either focusing on
the signal mass resolutions or the sensitivity. It is possible to utilise more
fundamental quantities, such as the charged particles or the calorimeter
deposits, to explore colour connections~\cite{colorqcd,colorbsm,rapiditygap}, but the showering and detector dependence have to be evaluated. 

The LHC will conclude Run-3 data-taking in the near future, and eventually, we
have to embrace the HL-LHC era. To achieve the ultimate sensitivity at the
HL-LHC, we need to maximise the discovery potential. FSR tagging offers a way
to further enhance the sensitivities, and it can be embedded in a multi-class
categorisation task to satisfy various analysis goals. The di-jet resonance
search has been the flagship inclusive search in hadron colliders. Despite its
long history, there is still space to enhance its potential, in particular with
modern ML-based technologies. The idea explored in this work may be
extended to other topics in the hadronic final states. We look forward to
seeing such techniques tested by the experiments. 

\section*{{Acknowledgments}}
We thank Sascha Dreyer, Christian Sander, Ryo Ishikawa and Yohei Yamaguchi for very constructive discussions. We also thank Marco Montella and Antonio Boveia for pointing relevant references to us. As undergraduates, Y.X. Shen and Y.S.Z. Sui contributed to sample generation, kinematic studies and the algorithm training/testing for this work substantially.  
\appendix
\section{Applications in Various Scenarios}
\label{app:app}
\subsection{Higher Mass Points}
\label{app:high_mass}

\begin{figure}[ht]
  \begin{center}
    \includegraphics[width=0.5\columnwidth]{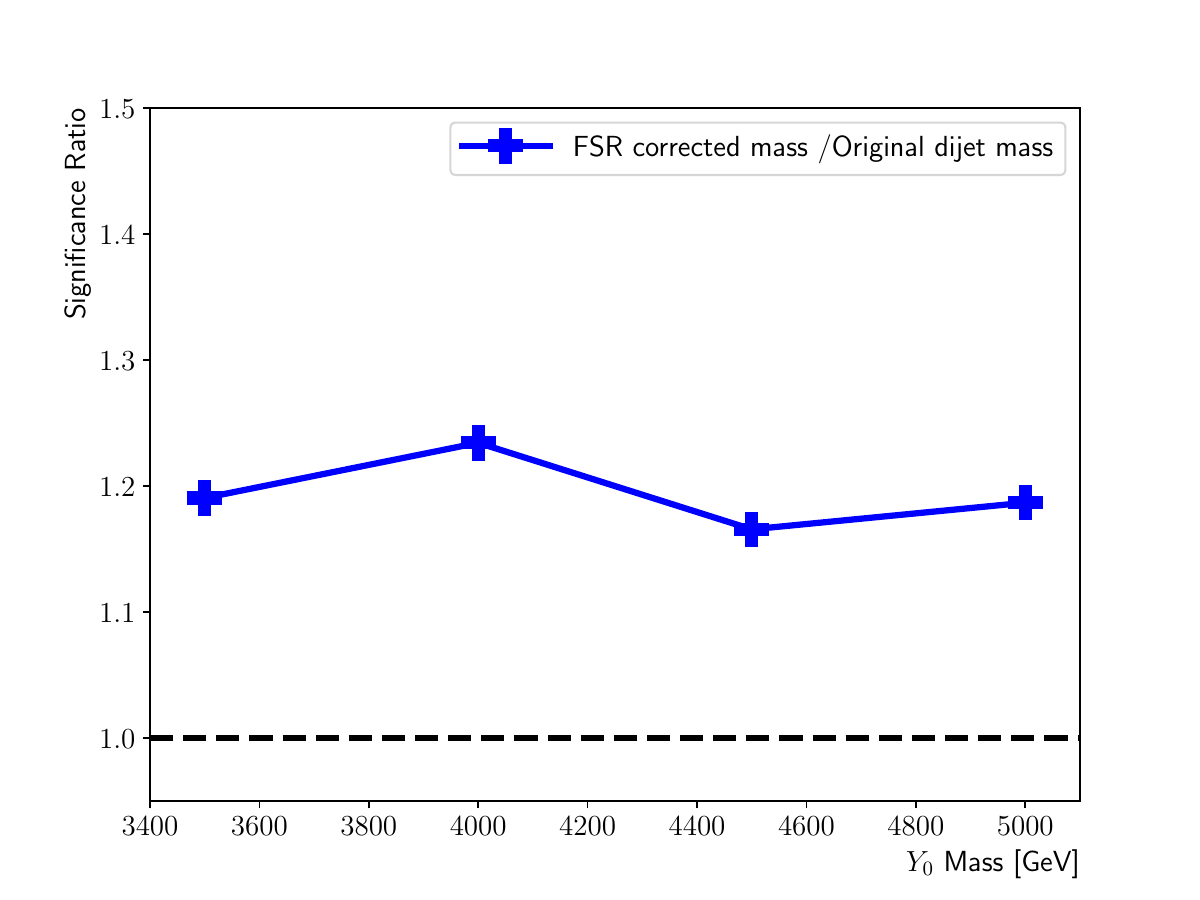}
    \caption{Summary of the ratios between $\max(\frac{\mathrm{N^i_{s}}}{\sqrt{\mathrm{N^i_B}}})$ obtained using the FSR corrected mass and that using the original di-jet mass, for the \mYZero = 3500 \GeV, 4000 \GeV, 4500 \GeV and 5000 \GeV signal points.}
    \label{fig:sen_com_ext}
  \end{center}
\end{figure}

\begin{figure}[ht]
  \begin{center}
    \includegraphics[width=0.45\columnwidth]{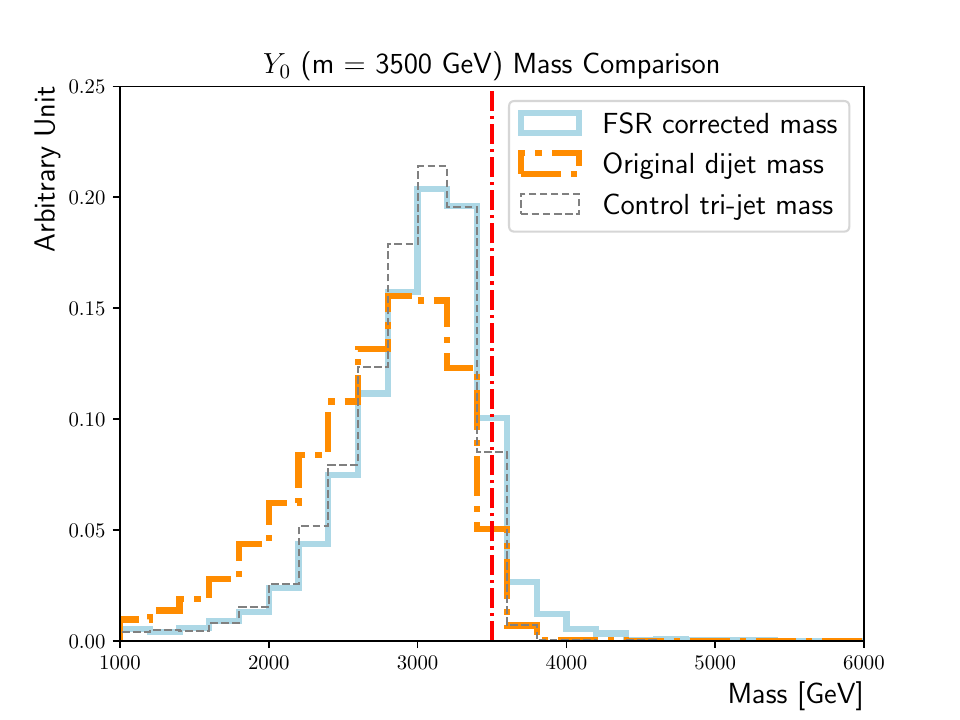}
    \includegraphics[width=0.45\columnwidth]{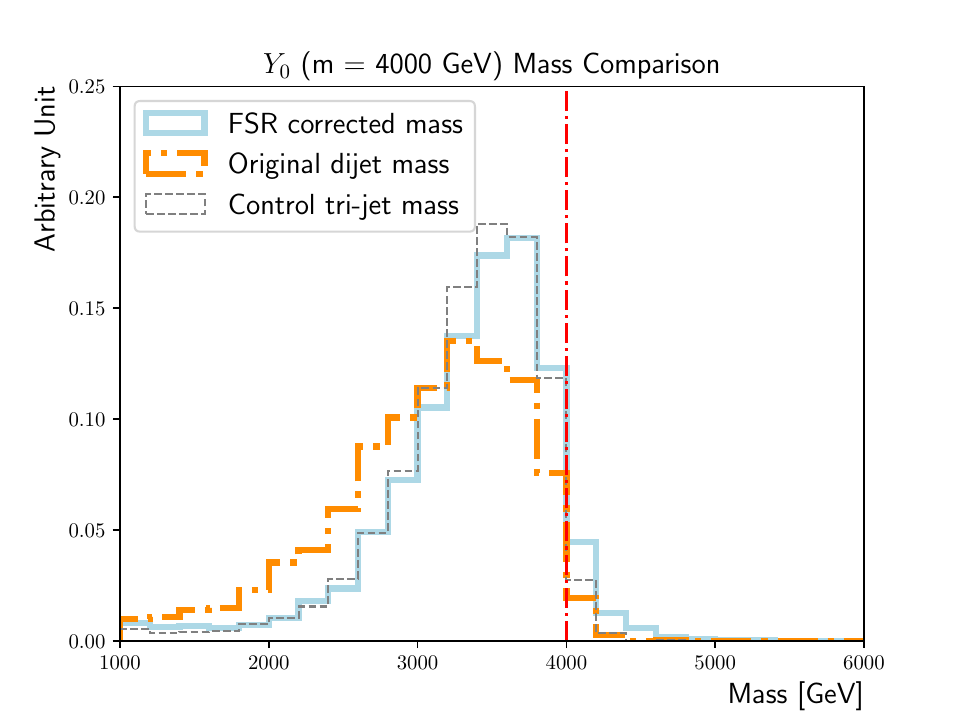}
    \includegraphics[width=0.45\columnwidth]{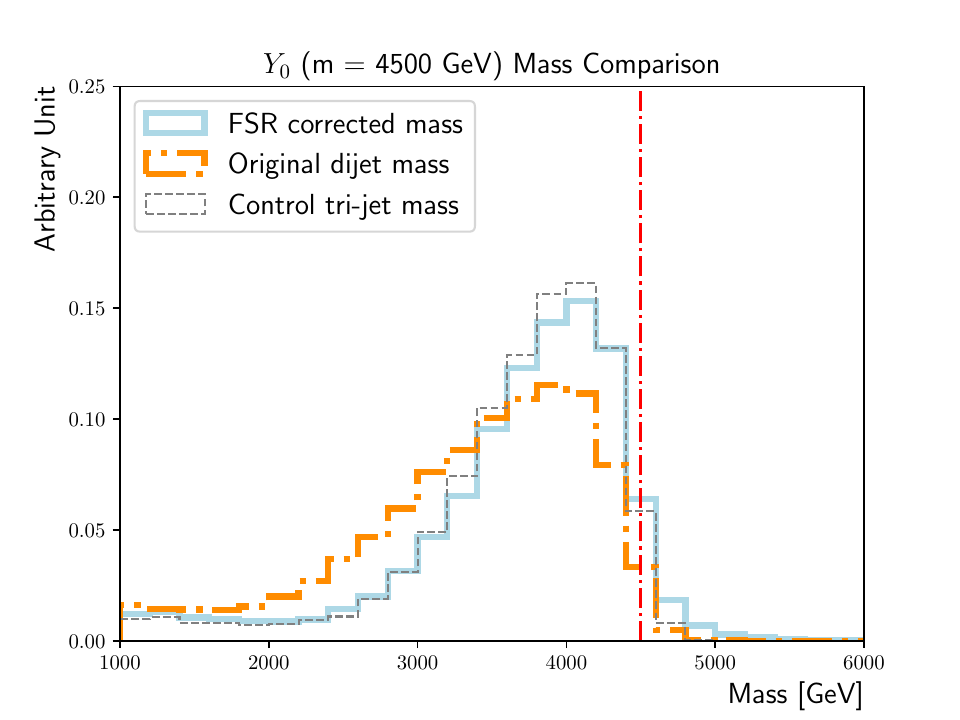}
    \includegraphics[width=0.45\columnwidth]{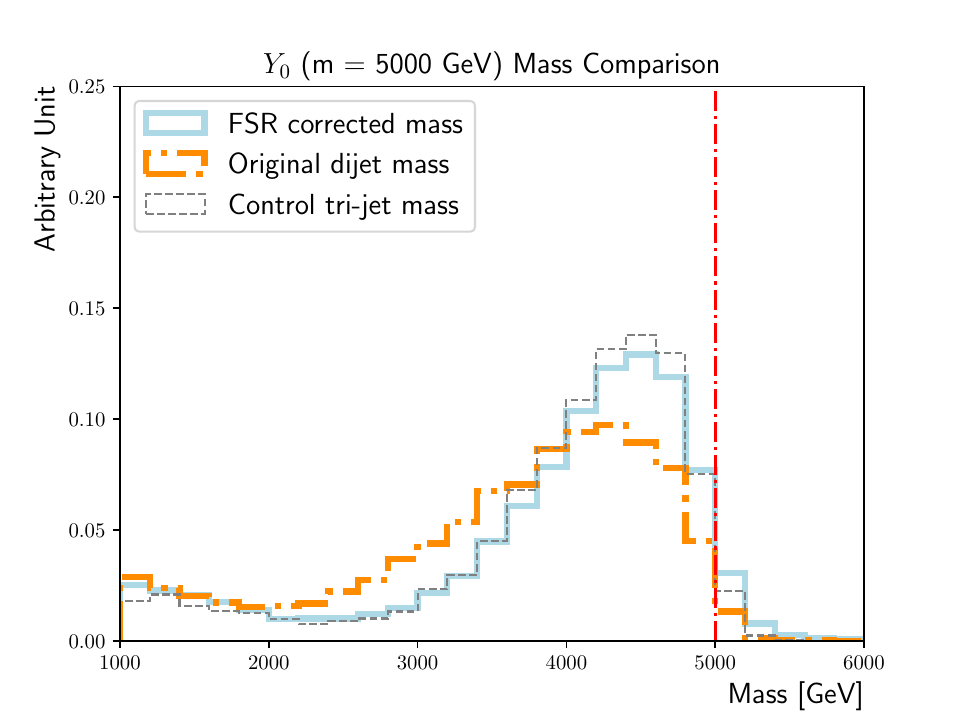}
    \caption{Comparisons of the FSR corrected masses (solid line) and the original di-jet masses (dotted-dashed line), for the \mYZero = 3500 \GeV (upper left), 4000 \GeV (upper right), 4500 \GeV (lower left) and 5000 \GeV (lower right) nominal signal samples. The tri-jet mass calculated using the showering control samples with the FSR/ISR showering switch turned on/off is added as a reference (dashed line). The actual \mYZero is indicated by the vertical line.}
    \label{fig:mass_com_sig_ext}
  \end{center}
\end{figure}

\begin{figure}[ht]
  \begin{center}
    \includegraphics[width=0.5\columnwidth]{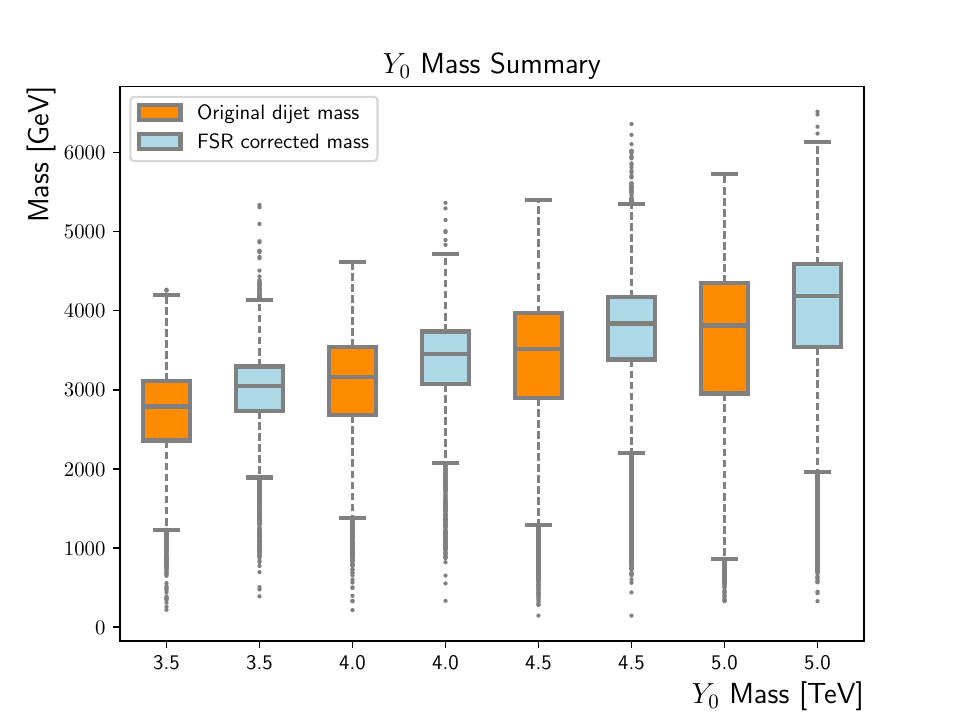}
    \caption{Summary of the FSR corrected mass distributions (light blue) and the original di-jet masses (dark orange), for \mYZero = 3500 \GeV, 4000 \GeV, 4500 \GeV and 5000 \GeV. The upper/lower boundary of the box indicates the 75\%/25\% percentile, while the centre line represents the median. The upper/lower error bar corresponds to the 95\%/5\% percentile.}
    \label{fig:mass_box_sig_ext}
  \end{center}
\end{figure}

The generality of the classifier is assessed using signal points, with \mYZero
ranging from 3.5 to 5 \TeV, that are not included in the training. Since
variables strongly correlated with the mass are excluded from the training, and
the training datasets are sampled to evenly populated in jet \pT, the
classifier brings similar sensitivity gains and mass resolution improvements in
the high mass region as well, seen in
Figure~\cref{fig:sen_com_ext,fig:mass_com_sig_ext,fig:mass_box_sig_ext}.   

\subsection{Re-defining the Discriminant}
\label{app:disc}

\begin{figure}[ht]
  \begin{center}
    \includegraphics[width=0.45\columnwidth]{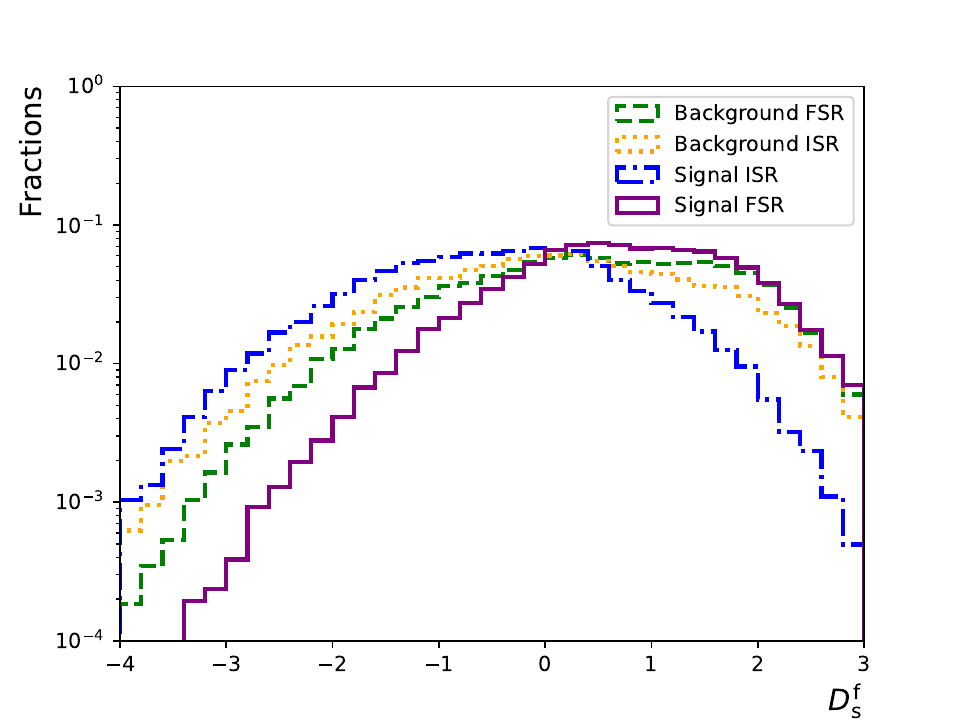}
    \includegraphics[width=0.45\columnwidth]{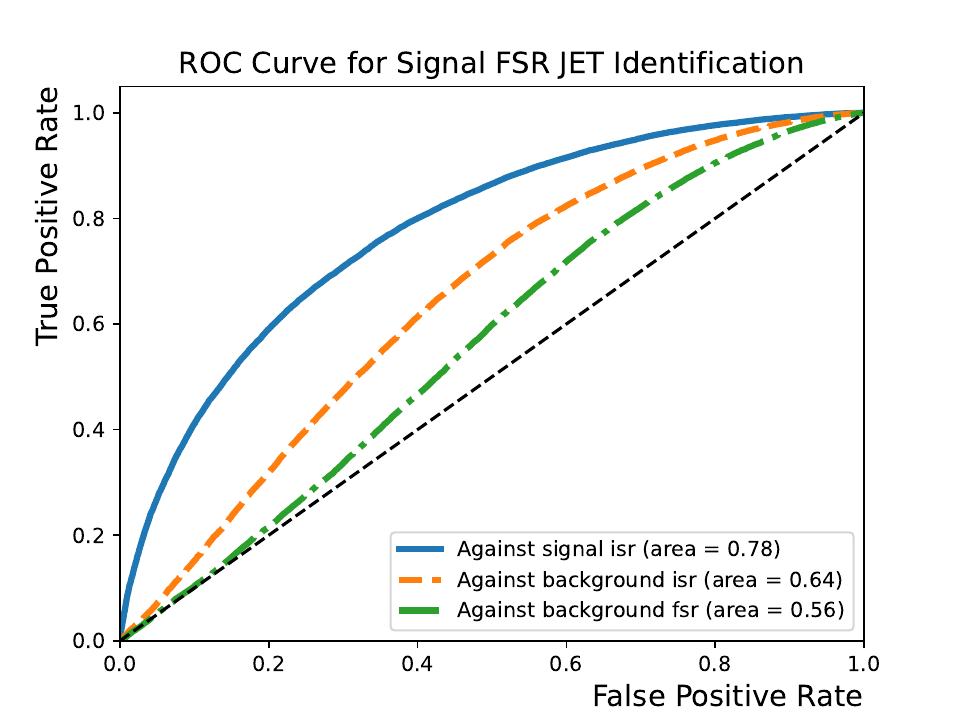}
    \caption{Left: Distributions of the $D_{\mathrm{s}}^{\mathrm{f}}$ for the ``sig-isr'' (dotted-dashed line), ``sig-fsr'' (solid line), ``bkg-isr'' (dotted line) and ``bkg-fsr'' (dashed line) categories. Right: ``sig-fsr'' identification efficiency as functions of the corresponding false positive rates for the ``sig-isr'' (solid line), ``bkg-isr'' (dashed line) and ``bkg-fsr'' (dotted-dashed line) categories. They are evaluated using the test dataset that accounts for 20\% of the total combined dataset. $f_{\mathrm{s}}^{\mathrm{i}}$ ($f_{\mathrm{b}}^{\mathrm{f}}$) are set to 1 (0).}
    \label{fig:rocs_alt}
  \end{center}
\end{figure}

The relative fractions, $f_{\mathrm{s}}^{\mathrm{i}}$ and
$f_{\mathrm{b}}^{\mathrm{f}}$, in the definition of
$D_{\mathrm{s}}^{\mathrm{f}}$, can be re-tuned for alternative use cases. In
Section~\ref{sec:perf}, they were chosen so that the classifier achieves similar
false positive rates for all the three categories, besides ``sig-fsr''. This
choice is motivated by the requirement that the background mass spectrum should
not be altered significantly. In a hypothetical case where such constraints are
mitigated, either due to advanced background modelling techniques or analysis
strategies, the classifier can be made fully concentrated on distinguishing
``sig-fsr'' from ``sig-isr'', by setting $f_{\mathrm{s}}^{\mathrm{i}}$
($f_{\mathrm{b}}^{\mathrm{f}}$) to 1 (0). Figure~\ref{fig:rocs_alt} shows the
corresponding $D_{\mathrm{s}}^{\mathrm{f}}$ distributions and the ROC curves.
Clearly, the rejection against ``sig-isr'' is significantly enhanced compared
to that in Figure~\ref{fig:rocs}.     

\clearpage
\addcontentsline{toc}{section}{References}
\bibliographystyle{CPC}
\bibliography{fsr}

@article{ISR2011,
    author = "Krohn, David and Randall, Lisa and Wang, Lian-Tao",
    title = "{On the Feasibility and Utility of ISR Tagging}",
    eprint = "1101.0810",
    archivePrefix = "arXiv",
    primaryClass = "hep-ph",
    month = "1",
    year = "2011"
}

@article{UA2,
    author = "{J. Alitti et al. (UA2 Collaboration)}",
    title = "{A Search for new intermediate vector mesons and excited quarks decaying to two jets at the CERN $\bar{p} p$ collider}",
    reportNumber = "CERN-PPE-93-66",
    doi = "10.1016/0550-3213(93)90395-6",
    journal = "Nucl. Phys. B",
    volume = "400",
    pages = "3--24",
    year = "1993"
}

@article{CDF,
    author = "{T. Aaltonen et al. (CDF Collaboration)}",
    title = "{Search for new particles decaying into dijets in proton-antiproton collisions at s**(1/2) = 1.96-TeV}",
    eprint = "0812.4036",
    archivePrefix = "arXiv",
    primaryClass = "hep-ex",
    reportNumber = "FERMILAB-PUB-08-572-E",
    doi = "10.1103/PhysRevD.79.112002",
    journal = "Phys. Rev. D",
    volume = "79",
    pages = "112002",
    year = "2009"
}

@article{D0,
    author = "{V.M. Abazov et al. (D0 Collaboration)}",
    title = "{Measurement of dijet angular distributions at s**(1/2) = 1.96-TeV and searches for quark compositeness and extra spatial dimensions}",
    eprint = "0906.4819",
    archivePrefix = "arXiv",
    primaryClass = "hep-ex",
    reportNumber = "FERMILAB-PUB-09-326-E",
    doi = "10.1103/PhysRevLett.103.191803",
    journal = "Phys. Rev. Lett.",
    volume = "103",
    pages = "191803",
    year = "2009"
}

@article{UA1,
    author = "{C. Albajar et al. (UA1 Collaboration)}",
    title = "{Two Jet Mass Distributions at the CERN Proton - Anti-Proton Collider}",
    reportNumber = "CERN-EP-88-54",
    doi = "10.1016/0370-2693(88)91843-6",
    journal = "Phys. Lett. B",
    volume = "209",
    pages = "127--134",
    year = "1988"
}

@article{ATLASDijet,
    author = "{ G. Aad et al. (ATLAS Collaboration)}",
    title = "{Search for new resonances in mass distributions of jet pairs using 139 fb$^{-1}$ of $pp$ collisions at $\sqrt{s}=13$ TeV with the ATLAS detector}",
    eprint = "1910.08447",
    archivePrefix = "arXiv",
    primaryClass = "hep-ex",
    reportNumber = "CERN-EP-2019-162",
    doi = "10.1007/JHEP03(2020)145",
    journal = "JHEP",
    volume = "03",
    pages = "145",
    year = "2020"
}

@article{ATLASTLA,
    author = "{ G. Aad et al. (ATLAS Collaboration)}",
    title = "{Search for electroweak-scale dijet resonances using trigger-level analysis with the ATLAS detector in $132$ fb$^{-1}$ of $pp$ collisions at $\sqrt{s} = 13$ TeV}",
    eprint = "2509.01219",
    archivePrefix = "arXiv",
    primaryClass = "hep-ex",
    reportNumber = "CERN-EP-2025-194",
    doi = "10.1103/15p2-bkg8",
    journal = "Phys. Rev. D",
    volume = "112",
    pages = "092015",
    year = "2025"
}

@article{CMSDijetAD,
    author = "{V. Chekhovsky et al. (CMS Collaboration)}",
    title = "{Model-agnostic search for dijet resonances with anomalous jet substructure in proton-proton collisions at $\sqrt{s}$ = 13 TeV}",
    eprint = "2412.03747",
    archivePrefix = "arXiv",
    primaryClass = "hep-ex",
    reportNumber = "CMS-EXO-22-026, CERN-EP-2024-291",
    doi = "10.1088/1361-6633/add762",
    journal = "Rept. Prog. Phys.",
    volume = "88",
    number = "6",
    pages = "067802",
    year = "2025"
}

@article{CMSDijet,
    author = "{A.M. Sirunyan et al. (CMS Collaboration)}",
    title = "{Search for high mass dijet resonances with a new background prediction method in proton-proton collisions at $\sqrt{s} =$ 13 TeV}",
    eprint = "1911.03947",
    archivePrefix = "arXiv",
    primaryClass = "hep-ex",
    reportNumber = "CMS-EXO-19-012, CERN-EP-2019-222",
    doi = "10.1007/JHEP05(2020)033",
    journal = "JHEP",
    volume = "05",
    pages = "033",
    year = "2020"
}

@article{madgraph,
    author = "Alwall, J. and Frederix, R. and Frixione, S. and Hirschi, V. and Maltoni, F. and Mattelaer, O. and Shao, H. -S. and Stelzer, T. and Torrielli, P. and Zaro, M.",
    title = "{The automated computation of tree-level and next-to-leading order differential cross sections, and their matching to parton shower simulations}",
    eprint = "1405.0301",
    archivePrefix = "arXiv",
    primaryClass = "hep-ph",
    reportNumber = "CERN-PH-TH-2014-064, CP3-14-18, LPN14-066, MCNET-14-09, ZU-TH-14-14",
    doi = "10.1007/JHEP07(2014)079",
    journal = "JHEP",
    volume = "07",
    pages = "079",
    year = "2014"
}

@article{pythia,
    author = {Sj\"ostrand, Torbj\"orn and Ask, Stefan and Christiansen, Jesper R. and Corke, Richard and Desai, Nishita and Ilten, Philip and Mrenna, Stephen and Prestel, Stefan and Rasmussen, Christine O. and Skands, Peter Z.},
    title = "{An introduction to PYTHIA 8.2}",
    eprint = "1410.3012",
    archivePrefix = "arXiv",
    primaryClass = "hep-ph",
    reportNumber = "LU-TP-14-36, MCNET-14-22, CERN-PH-TH-2014-190, FERMILAB-PUB-14-316-CD, DESY-14-178, SLAC-PUB-16122",
    doi = "10.1016/j.cpc.2015.01.024",
    journal = "Comput. Phys. Commun.",
    volume = "191",
    pages = "159--177",
    year = "2015"
}

@article{delphes,
    author = "de Favereau, J. and Delaere, C. and Demin, P. and Giammanco, A. and Lema\^\i{}tre, V. and Mertens, A. and Selvaggi, M.",
    collaboration = "DELPHES 3",
    title = "{DELPHES 3, A modular framework for fast simulation of a generic collider experiment}",
    eprint = "1307.6346",
    archivePrefix = "arXiv",
    primaryClass = "hep-ex",
    doi = "10.1007/JHEP02(2014)057",
    journal = "JHEP",
    volume = "02",
    pages = "057",
    year = "2014"
}

@article{colorbsm,
    author = "Alwall, Johan and de Visscher, Simon and Maltoni, Fabio",
    title = "{QCD radiation in the production of heavy colored particles at the LHC}",
    eprint = "0810.5350",
    archivePrefix = "arXiv",
    primaryClass = "hep-ph",
    doi = "10.1088/1126-6708/2009/02/017",
    journal = "JHEP",
    volume = "02",
    pages = "017",
    year = "2009"
}

@article{rapiditygap,
    author = "Sung, Ilmo",
    title = "{Probing the Gauge Content of Heavy Resonances with Soft Radiation}",
    eprint = "0908.3688",
    archivePrefix = "arXiv",
    primaryClass = "hep-ph",
    reportNumber = "YITP-SB-09-24",
    doi = "10.1103/PhysRevD.80.094020",
    journal = "Phys. Rev. D",
    volume = "80",
    pages = "094020",
    year = "2009"
}

@article{heavyrescolor,
    author = "Han, Tao and Lewis, Ian M. and Liu, Hongkai and Liu, Zhen and Wang, Xing",
    title = "{A guide to diagnosing colored resonances at hadron colliders}",
    eprint = "2306.00079",
    archivePrefix = "arXiv",
    primaryClass = "hep-ph",
    doi = "10.1007/JHEP08(2023)173",
    journal = "JHEP",
    volume = "08",
    pages = "173",
    year = "2023"
}

@article{jetbinning,
    author = "Ebert, Markus A. and Liebler, Stefan and Moult, Ian and Stewart, Iain W. and Tackmann, Frank J. and Tackmann, Kerstin and Zeune, Lisa",
    title = "{Exploiting jet binning to identify the initial state of high-mass resonances}",
    eprint = "1605.06114",
    archivePrefix = "arXiv",
    primaryClass = "hep-ph",
    reportNumber = "DESY-16-086, MIT-CTP-4804, NIKHEF-2016-019",
    doi = "10.1103/PhysRevD.94.051901",
    journal = "Phys. Rev. D",
    volume = "94",
    number = "5",
    pages = "051901",
    year = "2016"
}

@article{qcdnewphys,
    author = "Papaefstathiou, Andreas and Webber, Bryan",
    title = "{Effects of QCD radiation on inclusive variables for determining the scale of new physics at hadron colliders}",
    eprint = "0903.2013",
    archivePrefix = "arXiv",
    primaryClass = "hep-ph",
    reportNumber = "CAVENDISH-HEP-09-02, CERN-PH-TH-2009-029, MCNET-09-05",
    doi = "10.1088/1126-6708/2009/06/069",
    journal = "JHEP",
    volume = "06",
    pages = "069",
    year = "2009"
}

@article{isrmt2,
    author = "Nojiri, Mihoko M. and Sakurai, Kazuki",
    title = "{Controlling ISR in sparticle mass reconstruction}",
    eprint = "1008.1813",
    archivePrefix = "arXiv",
    primaryClass = "hep-ph",
    reportNumber = "KEK-TH-1388, CAVENDISH-HEP-10-13, DAMTP-2010-57",
    doi = "10.1103/PhysRevD.82.115026",
    journal = "Phys. Rev. D",
    volume = "82",
    pages = "115026",
    year = "2010"
}

@article{colorqcd,
    author = "Odagiri, Kosuke",
    title = "{Color connection structure of supersymmetric QCD (2 $\to$ 2) processes}",
    eprint = "hep-ph/9806531",
    archivePrefix = "arXiv",
    reportNumber = "CAVENDISH-HEP-98-05",
    doi = "10.1088/1126-6708/1998/10/006",
    journal = "JHEP",
    volume = "10",
    pages = "006",
    year = "1998"
}

@article{dijetreview,
    author = "Harris, Robert M. and Kousouris, Konstantinos",
    title = "{Searches for Dijet Resonances at Hadron Colliders}",
    eprint = "1110.5302",
    archivePrefix = "arXiv",
    primaryClass = "hep-ex",
    reportNumber = "FERMILAB-PUB-11-567, CERN-OPEN-2011-044",
    doi = "10.1142/S0217751X11054905",
    journal = "Int. J. Mod. Phys. A",
    volume = "26",
    pages = "5005--5055",
    year = "2011"
}

@article{atlasexo,
    author = "{ G. Aad et al. (ATLAS Collaboration)}",   
     title = "{Exploration at the high-energy frontier: ATLAS Run 2 searches investigating the exotic jungle beyond the Standard Model}",
     eprint = "2403.09292",
     archivePrefix = "arXiv",
     primaryClass = "hep-ex",
     reportNumber = "CERN-EP-2024-075",
     doi = "10.1016/j.physrep.2024.10.001",
     journal = "Phys. Rept.",
     volume = "1116",
     pages = "301--385",
     year = "2025"
}

@article{tlareview,
     author = "{A. Hayrapetyan et al. (CMS Collaboration)}",
     title = "{Enriching the physics program of the CMS experiment via data scouting and data parking}",
     eprint = "2403.16134",
     archivePrefix = "arXiv",
     primaryClass = "hep-ex",
     reportNumber = "CMS-EXO-23-007, CERN-EP-2024-068",
     doi = "10.1016/j.physrep.2024.09.006",
     journal = "Phys. Rept.",
     volume = "1115",
     pages = "678--772",
     year = "2025"
}

@article{atlasdijetisr,
    author = "{ G. Aad et al. (ATLAS Collaboration)}",
    title = "{Search for low-mass resonances decaying into two jets and produced in association with a photon or a jet at $\sqrt{s}$=13 TeV with the ATLAS detector}",
    eprint = "2403.08547",
    archivePrefix = "arXiv",
    primaryClass = "hep-ex",
    reportNumber = "CERN-EP-2024-044",
    doi = "10.1103/PhysRevD.110.032002",
    journal = "Phys. Rev. D",
    volume = "110",
    number = "3",
    pages = "032002",
    year = "2024"
}

@article{cmsdijetisr,
    author = "{A.M. Sirunyan (CMS Collaboration)}",
    title = "{Search for dijet resonances using events with three jets in proton-proton collisions at s=13TeV}",
    eprint = "1911.03761",
    archivePrefix = "arXiv",
    primaryClass = "hep-ex",
    reportNumber = "CMS-EXO-19-004, CERN-EP-2019-229",
    doi = "10.1016/j.physletb.2020.135448",
    journal = "Phys. Lett. B",
    volume = "805",
    pages = "135448",
    year = "2020"
}

@article{pythiacode,
    author = "Bierlich, Christian and others",
    title = "{A comprehensive guide to the physics and usage of PYTHIA 8.3}",
    eprint = "2203.11601",
    archivePrefix = "arXiv",
    primaryClass = "hep-ph",
    reportNumber = "LU-TP 22-16, MCNET-22-04, FERMILAB-PUB-22-227-SCD",
    doi = "10.21468/SciPostPhysCodeb.8",
    journal = "SciPost Phys. Codeb.",
    volume = "2022",
    pages = "8",
    year = "2022"
}

@article{dm_simp1,
    author = {Backovi{\'c}, Mihailo and Kr{\"a}mer, Michael and Maltoni, Fabio and Martini, Antony and Mawatari, Kentarou and Pellen, Mathieu},
    title = "{Higher-order QCD predictions for dark matter production at the LHC in simplified models with s-channel mediators}",
    eprint = "1508.05327",
    archivePrefix = "arXiv",
    primaryClass = "hep-ph",
    reportNumber = "MCNET-15-24, CP3-15-25, TTK-15-19",
    doi = "10.1140/epjc/s10052-015-3700-6",
    journal = "Eur. Phys. J. C",
    volume = "75",
    number = "10",
    pages = "482",
    year = "2015"
}

@article{dm_simp,
    author = "Mattelaer, Olivier and Vryonidou, Eleni",
    title = "{Dark matter production through loop-induced processes at the LHC: the s-channel mediator case}",
    eprint = "1508.00564",
    archivePrefix = "arXiv",
    primaryClass = "hep-ph",
    reportNumber = "IPPP-15-48, DCPT-15-96, CP3-15-23, MCNET-15-20",
    doi = "10.1140/epjc/s10052-015-3665-5",
    journal = "Eur. Phys. J. C",
    volume = "75",
    number = "9",
    pages = "436",
    year = "2015"
}

@article{dm_simp2,
    author = "Albert, Andreas and others",
    title = "{Recommendations of the LHC Dark Matter Working Group: Comparing LHC searches for dark matter mediators in visible and invisible decay channels and calculations of the thermal relic density}",
    eprint = "1703.05703",
    archivePrefix = "arXiv",
    primaryClass = "hep-ex",
    reportNumber = "CERN-LPCC-2017-01",
    doi = "10.1016/j.dark.2019.100377",
    journal = "Phys. Dark Univ.",
    volume = "26",
    pages = "100377",
    year = "2019"
}

@article{excitedquark1,
author = {Baur, U. and Hinchliffe, I. and Zeppenfeld, D.},
title = {Excited quark production at hadron colliders},
journal = {International Journal of Modern Physics A},
volume = {02},
number = {04},
pages = {1285-1297},
year = {1987},
doi = {10.1142/S0217751X87000661},
}

@article{excitedquark2,
  title = {Excited-quark and -lepton production at hadron colliders},
  author = {Baur, U. and Spira, M. and Zerwas, P. M.},
  journal = {Phys. Rev. D},
  volume = {42},
  issue = {3},
  pages = {815--824},
  year = {1990},
  month = {Aug},
  publisher = {American Physical Society},
  doi = {10.1103/PhysRevD.42.815},
  url = {https://link.aps.org/doi/10.1103/PhysRevD.42.815}
}

@article{extradimension,
    author = "Cullen, Schuyler and Perelstein, Maxim and Peskin, Michael E.",
    title = "{TeV strings and collider probes of large extra dimensions}",
    eprint = "hep-ph/0001166",
    archivePrefix = "arXiv",
    reportNumber = "SLAC-PUB-8319, SU-ITP-99-53",
    doi = "10.1103/PhysRevD.62.055012",
    journal = "Phys. Rev. D",
    volume = "62",
    pages = "055012",
    year = "2000"
}

@article{qbh,
    author = "Anchordoqui, Luis A. and Feng, Jonathan L. and Goldberg, Haim and Shapere, Alfred D.",
    title = "{Inelastic black hole production and large extra dimensions}",
    eprint = "hep-ph/0311365",
    archivePrefix = "arXiv",
    reportNumber = "NUB-3243-TH-03, UCI-TR-2003-30, UK-03-15",
    doi = "10.1016/j.physletb.2004.05.051",
    journal = "Phys. Lett. B",
    volume = "594",
    pages = "363--367",
    year = "2004"
}

@article{qgtag,
    author = "{ G. Aad et al. (ATLAS Collaboration)}",
    title = "{Performance and calibration of quark/gluon-jet taggers using 140 fb$^{-1}$ of pp collisions at $\sqrt{s}=13$ TeV with the ATLAS detector}",
    eprint = "2308.00716",
    archivePrefix = "arXiv",
    primaryClass = "hep-ex",
    reportNumber = "CERN-EP-2023-151",
    doi = "10.1088/1674-1137/acf701",
    journal = "Chin. Phys. C",
    volume = "48",
    number = "2",
    pages = "023001",
    year = "2024"
}

@article{abcd,
    author = "Boveia, Antonio and Doglioni, Caterina",
    title = "{Dark Matter Searches at Colliders}",
    eprint = "1810.12238",
    archivePrefix = "arXiv",
    primaryClass = "hep-ex",
    doi = "10.1146/annurev-nucl-101917-021008",
    journal = "Ann. Rev. Nucl. Part. Sci.",
    volume = "68",
    pages = "429--459",
    year = "2018"
}

@article{svj,
    author = "Cohen, Timothy and Lisanti, Mariangela and Lou, Hou Keong",
    title = "{Semivisible Jets: Dark Matter Undercover at the LHC}",
    eprint = "1503.00009",
    archivePrefix = "arXiv",
    primaryClass = "hep-ph",
    doi = "10.1103/PhysRevLett.115.171804",
    journal = "Phys. Rev. Lett.",
    volume = "115",
    number = "17",
    pages = "171804",
    year = "2015"
}

@article{emj,
    author = "Schwaller, Pedro and Stolarski, Daniel and Weiler, Andreas",
    title = "{Emerging Jets}",
    eprint = "1502.05409",
    archivePrefix = "arXiv",
    primaryClass = "hep-ph",
    reportNumber = "CERN-PH-TH-2015-031, DESY-15-026",
    doi = "10.1007/JHEP05(2015)059",
    journal = "JHEP",
    volume = "05",
    pages = "059",
    year = "2015"
}

@article{jvt,
    author = "M. Aaboud et al. (ATLAS Collaboration)", 
    title = "{Identification and rejection of pile-up jets at high pseudorapidity with the ATLAS detector}",
    eprint = "1705.02211",
    archivePrefix = "arXiv",
    primaryClass = "hep-ex",
    reportNumber = "CERN-EP-2017-055",
    doi = "10.1140/epjc/s10052-017-5081-5",
    journal = "Eur. Phys. J. C",
    volume = "77",
    number = "9",
    pages = "580",
    year = "2017",
    note = "[Erratum: Eur.Phys.J.C 77, 712 (2017)]"
}

@article{bingsvj,
    author = "Liu, Bingxuan and Pedro, Kevin",
    title = "{Semi-visible jets + X: illuminating dark showers with radiation}",
    eprint = "2409.04741",
    archivePrefix = "arXiv",
    primaryClass = "hep-ph",
    reportNumber = "FERMILAB-PUB-24-0563-CSAID-PPD",
    doi = "10.1007/JHEP12(2024)105",
    journal = "JHEP",
    volume = "12",
    pages = "105",
    year = "2024"
}

@article{cmssvj,
    author = "{A. Tumasyan et al. (CMS Collaboration)}",
    title = "{Search for resonant production of strongly coupled dark matter in proton-proton collisions at 13 TeV}",
    eprint = "2112.11125",
    archivePrefix = "arXiv",
    primaryClass = "hep-ex",
    reportNumber = "CMS-EXO-19-020, CERN-EP-2021-252",
    doi = "10.1007/JHEP06(2022)156",
    journal = "JHEP",
    volume = "06",
    pages = "156",
    year = "2022"
}

@article{antikt,
    author = "Cacciari, Matteo and Salam, Gavin P. and Soyez, Gregory",
    title = "{The anti-$k_t$ jet clustering algorithm}",
    eprint = "0802.1189",
    archivePrefix = "arXiv",
    primaryClass = "hep-ph",
    reportNumber = "LPTHE-07-03",
    doi = "10.1088/1126-6708/2008/04/063",
    journal = "JHEP",
    volume = "04",
    pages = "063",
    year = "2008"
}

@article{catchment,
    author = "Cacciari, Matteo and Salam, Gavin P. and Soyez, Gregory",
    title = "{The Catchment Area of Jets}",
    eprint = "0802.1188",
    archivePrefix = "arXiv",
    primaryClass = "hep-ph",
    reportNumber = "LPTHE-07-02",
    doi = "10.1088/1126-6708/2008/04/005",
    journal = "JHEP",
    volume = "04",
    pages = "005",
    year = "2008"
}

@article{GN2,
    author = "{ G. Aad et al. (ATLAS Collaboration)}",
    title = "{Transforming jet flavour tagging at ATLAS}",
    eprint = "2505.19689",
    archivePrefix = "arXiv",
    primaryClass = "hep-ex",
    reportNumber = "CERN-EP-2025-103",
    month = "5",
    year = "2025"
}

@article{FD,
    author = "Edgar, Ryan and Amidei, Dante and Grud, Christopher and Sekhon, Karishma",
    title = "{Functional Decomposition: A new method for search and limit setting}",
    eprint = "1805.04536",
    archivePrefix = "arXiv",
    primaryClass = "physics.data-an",
    month = "5",
    year = "2018"
}

@article{GPR3,
    author = "Gandrakota, Abhijith and Lath, Amit and Morozov, Alexandre V. and Murthy, Sindhu",
    title = "{Model selection and signal extraction using Gaussian Process regression}",
    eprint = "2202.05856",
    archivePrefix = "arXiv",
    primaryClass = "hep-ex",
    reportNumber = "FERMILAB-PUB-22-073-CMS",
    doi = "10.1007/JHEP02(2023)230",
    journal = "JHEP",
    volume = "02",
    pages = "230",
    year = "2023"
}

@book{GPR1,
  author = {Rasmussen, Carl Edward and Williams, Christopher K. I.},
  publisher = {The MIT Press},
  title = {Gaussian Processes for Machine Learning},
  year = 2006
}

@article{GPR2,
    author = "Frate, Meghan and Cranmer, Kyle and Kalia, Saarik and Vandenberg-Rodes, Alexander and Whiteson, Daniel",
    title = "{Modeling Smooth Backgrounds and Generic Localized Signals with Gaussian Processes}",
    eprint = "1709.05681",
    archivePrefix = "arXiv",
    primaryClass = "physics.data-an",
    month = "9",
    year = "2017"
}

@article{GPR4,
    author = "Barr, Jackson and Liu, Bingxuan",
    title = "{Gaussian process regression as a sustainable data-driven background estimate method at the (HL)-LHC}",
    eprint = "2503.07289",
    archivePrefix = "arXiv",
    primaryClass = "hep-ex",
    doi = "10.1140/epjc/s10052-025-14574-3",
    journal = "Eur. Phys. J. C",
    volume = "85",
    number = "8",
    pages = "846",
    year = "2025"
}

@article{SymbolFit,
    author = "Tsoi, Ho Fung and Rankin, Dylan and Caillol, Cecile and Cranmer, Miles and Dasu, Sridhara and Duarte, Javier and Harris, Philip and Lipeles, Elliot and Loncar, Vladimir",
    title = "{SymbolFit: Automatic Parametric Modeling with Symbolic Regression}",
    eprint = "2411.09851",
    archivePrefix = "arXiv",
    primaryClass = "hep-ex",
    doi = "10.1007/s41781-025-00140-9",
    journal = "Comput. Softw. Big Sci.",
    volume = "9",
    number = "1",
    pages = "12",
    year = "2025"
}

@article{inter1,
    author = "Bhattiprolu, Prudhvi N. and Martin, Stephen P.",
    title = "{Signal-background interference for digluon resonances at the Large Hadron Collider}",
    eprint = "2004.06181",
    archivePrefix = "arXiv",
    primaryClass = "hep-ph",
    doi = "10.1103/PhysRevD.102.015016",
    journal = "Phys. Rev. D",
    volume = "102",
    number = "1",
    pages = "015016",
    year = "2020"
}

@article{inter2,
    author = "Martin, Stephen P.",
    title = "{Signal-background interference for a singlet spin-0 digluon resonance at the LHC}",
    eprint = "1606.03026",
    archivePrefix = "arXiv",
    primaryClass = "hep-ph",
    doi = "10.1103/PhysRevD.94.035003",
    journal = "Phys. Rev. D",
    volume = "94",
    number = "3",
    pages = "035003",
    year = "2016"
}

@article{inter3,
    author = "Bian, Ligong and Liu, Da and Shu, Jing and Zhang, Yongchao",
    title = "{Interference Effect on Resonance Studies in Searches of Heavy Particles}",
    eprint = "1509.02787",
    archivePrefix = "arXiv",
    primaryClass = "hep-ph",
    reportNumber = "ULB-TH-16-05",
    doi = "10.1142/S0217751X16500834",
    journal = "Int. J. Mod. Phys.",
    volume = "31",
    number = "14n15",
    pages = "1650083",
    year = "2016"
}

@article{relu,
    author={Fukushima, Kunihiko},
    title={Cognitron: A self-organizing multilayered neural network},
    journal={Biological Cybernetics},
    year={1975},
    month={9},
    day={01},
    volume={20},
    number={3},
    pages={121-136},
    doi={10.1007/BF00342633},
    url={https://doi.org/10.1007/BF00342633}
}

@article{SGD,
author = {Robbins, Herbert and Monro, Sutton},
copyright = {Copyright 1951 Institute of Mathematical Statistics},
issn = {0003-4851},
journal = {The Annals of mathematical statistics},
language = {eng},
number = {3},
pages = {400-407},
publisher = {Institute of Mathematical Statistics},
title = {A Stochastic Approximation Method},
volume = {22},
year = {1951},
}

\end{document}